\documentclass[11pt]{article}
\usepackage{amssymb}
\usepackage{graphicx}

[1999/03/29 PSNFSS v.7.2
Times font as default roman
: S Rahtz]

\newcommand{\be}{\begin{equation}}
\newcommand{\ee}{\end{equation}}
\newcommand{\bea}{\begin{eqnarray}}
\newcommand{\eea}{\end{eqnarray}}

\begin{document}

\title{{\bf General framework for a portfolio theory with
non-Gaussian risks and non-linear correlations}}

\author{\bf Y. Malevergne$^1$ and D. Sornette$^{1,2}$\\
$^1$ Laboratoire de Physique de la Mati\`{e}re Condens\'{e}e\\ CNRS UMR6622 and
Universit\'{e} de Nice-Sophia Antipolis\\ Parc
Valrose, 06108 Nice Cedex 2, France \\
$^2$ Institute of Geophysics and
Planetary Physics and Department of Earth and Space Science\\
University of California, Los Angeles, California 90095\\
e-mails: Yannick.Malevergne@unice.fr and sornette@unice.fr
}

\maketitle

\begin{abstract}
\noindent Using a family of modified Weibull distributions,
encompassing both sub-exponentials and super-exponentials, to parameterize
the marginal distributions of asset returns and their natural multivariate
generalizations, we give exact formulas for the tails and for the
moments and cumulants of the
distribution of returns of a portfolio make of arbitrary compositions of these
assets. Using combinatorial and hypergeometric functions,
we are in particular able to extend previous results to the case
where the exponents of the Weibull distributions are different from
asset to asset and
in the presence of dependence between assets. We treat in details the problem
of risk minimization using two different measures of risks (cumulants and
value-at-risk) for a portfolio made of two assets
and compare the theoretical predictions with direct empirical data.
While good agreement is found, the remaining discrepancy between
theory and data
stems from the deviations from the Weibull parameterization for small returns.
Our extended formulas enable us to determine analytically the
conditions under which
it is possible to ``have your cake and eat it too'', i.e.,
to construct a portfolio with both larger return and smaller ``large risks''.

\end{abstract}

\section{Introduction}

The determination of the risks and returns associated with a given
portfolio constituted
of $N$ assets is completely embedded in the knowledge of their
multivariate distribution
of returns. Indeed,
the dependence between random variables is completely described by
their joint distribution.
This remark entails the two major problems of portfolio theory: 1) determine
the multivariate distribution function of asset returns; 2) derive
from it useful measures of
portfolio risks and use them to analyze and optimize portfolios.
Here, we address them
both by extending the new approach of \cite{Sornette1} in terms of a
class of subexponential and superexponential multivariate distributions.

In the standard Gaussian framework, the multivariate distribution
takes the form
of an exponential of minus a quadratic form $X' \Omega^{-1} X$, where
$X$ is the unicolumn
of asset returns and $\Omega$ is their covariance matrix. The beauty
and simplicity
of the Gaussian case is that the essentially impossible task
of determining a large multidimensional function is collapsed into
the very much simpler
one of calculating the $N(N+1)/2$ elements of the symmetric covariance matrix.
Risk is then uniquely and completely embodied by the variance of the
portfolio return,
which is easily determined from the covariance matrix. This is the basis of
Markovitz's portfolio theory \cite{Markovitz} and of the CAPM (see
for instance \cite{Merton}).
In this framework, increasing return comes with risk and return may
remunerate a larger risk.

However, the variance (volatility) of portfolio returns provides at
best a limited
quantification of incurred risks, as the empirical distributions of
returns have ``fat tails''
\cite{Lux,Gopikrishnan} and
the dependences between assets are only imperfectly accounted for by
the covariance
matrix \cite{Litterman}. Value-at-Risk \cite{Jorion}
and other measures of risks \cite{Artzner,Sornettepre,Bouchaudetal,Sornette1}
have been developed to account for the larger moves allowed by
non-Gaussian distributions
and nonlinear correlations.

In section 2, we present our parameterization of the multivariate
distribution of
returns based on two steps: (i) the projection of the empirical
marginal distributions
onto Gaussian laws via nonlinear mappings; (ii) the use of an entropy
maximization
to construct the corresponding most parsimonious
representation of the multivariate distribution. We show in
particular that this
construction amounts to use Gaussian copulas. Empirical tests are
given which show that
this assumption is a very good approxition.

Section 3 offers a specific parameterization
of marginal distributions in terms of so-called modified Weibull
distributions, which are
essentially exponential of minus a power law. This family of
distribution contains both
sub-exponential and super-exponentials, inclusing the Gaussian law as
a special case.
Notwithstanding their possible fat-tail nature, all their moments and cumulants
are finite and can be calculated.
We present empirical calibration of the two key parameters of the
modified Weibull
distribution, namely the exponent $c$ and the characteristic scale $\chi$.

Section 4 uses the multivariate construction based on (i) the modified Weibull
marginal distributions and (ii) the Gaussian copula to derive the
asymptotic analytical
form of the tail of the distribution of returns of a portfolio composed of an
arbitrary combination of these assets. We show that, in the case where
individual asset returns have
the asymptotic tail $\exp[-(x_i/\chi_i)^c]$ with the same exponent $c$, then
the tail of the distribution of portfolio return $S$ is of the same form
$\exp[-(S/\chi)^c]$, with the same exponent $c$ but with a
characteristic scale $\chi$
taking different functional forms depending on the value of $c$ and
the strength
of the dependence between the assets. $\chi$ is also of course a function
of the asset weights in the portfolio.
These results allow one to estimate the value-at-risk (VaR)
in this non-Gaussian nonlinear dependence framework.

Section 5 provides the analytical expressions of the cumulants of the
distribution of
portfolio returns. Cumulants are of interest because
they are natural measures of risks, the higher their order, the
more weight being given to large risks. Recall that the second
cumulant is nothing but
the variance, the normalized third-order (resp. fourth-order)
cumulant is the skewness (resp. excess kurtosis). This section
provides the most
general formulas for any possible positive values of the exponents
$c$ of the asset
return distributions, generalizing generously previous results
\cite{Sornette1}.
In the case of dependent assets, we give the rather cumbersome
explicit formulas
only for the case of portfolios of two assets. Similar more
cumbersome expressions hold for the general case of $N$ assets. This
section also
offers empirical tests comparing the direct numerical evaluation of
the cumulants
of financial time series to the values predicted from our analytical formulas
using the exponents $c$ and characteristic scales $\chi$ calibrated
previously on the
same data. Good consistency is found.

Section 6 uses these two sets of results to offer a first approach to portfolio
optimization. We use two approaches, one based on the asymptotic form of the
tail of the distribution of portfolio returns, the other based on the
expression
of the cumulants of the distribution of the portfolio returns. In
both cases, we
show how to generalize the concept of an efficient frontier,
initially introduced
in the mean-variance space. We extend it using the different measures
of risks captured
by the tail of the distribution and by the cumulants. The main novel result
is an analytical understanding of the conditions under which it is possible to
simultaneously increase the portfolio return and decreases its large
risks quantified
by large-order cumulants. It thus appears that the multidimensional
nature of risks
allows one to break the stalemate of no better return without more risks.
Section 7 concludes.

Before proceeding with the presentation of our results, we set the notations to
derive the basic problem addressed in this paper, namely to study the
distribution
of the sum of weighted random variables with arbitrary marginal
distributions and
dependence.  Consider a portfolio with $n_i$ shares of asset
$i$ of price $p_i(0)$ at time $t=0$ whose initial wealth is
\be
W(0) = \sum_{i=1}^N n_i p_i(0)~.
\ee
A time $\tau$ later, the wealth has become $W(\tau) = \sum_{i=1}^N n_i
p_i(\tau)$ and the wealth variation is
\be
\delta_{\tau} W \equiv W(\tau) -W(0) = \sum_{i=1}^N n_i p_i(0)
{{p_i(\tau) - p_i(0)} \over p_i(0)}
= W(0) ~\sum_{i=1}^N w_i r_i(t,\tau) ,
\ee
where
\be
w_i = {n_i p_i(0)  \over \sum_{j=1}^{N} n_j p_j(0)}~
\ee
is the fraction in capital invested in the $i$th asset at time $0$ and
the return
$r_i(t,\tau)$ between time $t-\tau$ and $t$ of asset $i$ is defined as:
\be
 r_i(t,\tau) = {p_i(t)-p_i(t-\tau) \over p_i(t-\tau)} ~.
 \label{jhghss}
\ee
Using the definition (\ref{jhghss}),
this justifies us to write the return $S_{\tau}$ of the portfolio over a time
interval $\tau$ as the weighted sum of the returns $r_i(\tau)$ of the assets
$i=1,...,N$ over the time interval $\tau$
\be
S{\tau} = {\delta_{\tau} W \over W(0)} = \sum_{i=1}^N w_i ~r_i(\tau)~.
\label{jjkmmq}
\ee
In the sequel, we shall thus consider asset returns as the
fundamental variables
(denoted $x_i$ or $X_i$ in the sequel)
and study their aggregation properties, namely how the distribution
of portfolio
return equal to their weighted sum derives for their multivariable
distribution.
We shall consider a single time scale $\tau$ which can be chosen arbitrarily,
say equal to one day. We shall thus drop the dependence on $\tau$,
understanding
implicitely that all our results hold for returns estimated over time
step $\tau$.

\section{Estimation of the joint probability distribution of returns
of several assets}

\subsection{A brief exposition and justification of the method}

We will use the method of determination of multivariate distributions
introduced by Karlen \cite{Karlen} and Sornette et al. \cite{Sornette1}. This
method consists in two steps: (i) transform each return $x$ into a
Gaussian variable $y$ by a nonlinear monotonous increasing
mapping; (ii) use the principle of entropy
maximization to construct the corresponding multivariate distribution
of the transformed variables $y$.

The first concern to address before going any further
is whether the nonlinear transformation, which is
in principle different for each asset return, conserves the structure of the
dependence. In what sense is the dependence between the transformed
variables $y$
the same as the dependence between the asset returns $x$? It turns out
that the notion of ``copulas'' provides a general and rigorous answer which
justifies our procedure \cite{Sornette1}.

For completeness, we briefly recall the definition of a copula (for further
details about the concept of copula see \cite{Nelsen}).  A
function  $C$ : $[0,1]^n \longrightarrow [0,1]$ is a $n$-copula if it
enjoys the
following properties~:
\begin{itemize}
\item $\forall u \in [0,1] $, $C(1,\cdots, 1, u,
1 \cdots, 1)=u$~,
\item $\forall u_i \in [0,1] $, $C(u_1, \cdots, u_n)=0$ if
at least one of the $u_i$ equals zero~,
\item $C$ is grounded and $n$-increasing i.e the $C$-volume of every
boxes whose
vertices lie in $[0,1]^n$ is positive.
\end{itemize}

Skar's Theorem then states that, given an $n$-dimensional
distribution function  $F$ with continuous marginal distributions
$F_1, \cdots, F_n$, there
exists a unique $n$-copula $C$ : $[0,1]^n \longrightarrow [0,1]$ such
that~: \be
F(x_1, \cdots, x_n) = C(F_1(x_1), \cdots, F_n(x_n))~.
\ee
This elegant result shows that the study of the dependence of random variables
can be performed independently of the behavior of the marginal distributions.
Moreover, the following result shows that copulas are intrinsic measures of
dependence. Consider $n$ continous random variables  $X_1, \cdots ,
X_n$ with copula $C$.
Then, if $g_1(X_1), \cdots, g_n(X_n)$ are
strictly increasing on the ranges of $X_1, \cdots , X_n$, the random variables
$Y_1=g_1(X_1), \cdots, Y_n=g_n(X_n)$ have exactly the same copula $C$
\cite{Lindskog}.
The copula is
thus invariant under strictly increasing tranformation of the variables.
This provides a powerful way
of studying scale-invariant measures of associations.
It is also a natural starting point for construction of
multivariate distributions and provides the theorical
justification of the method of determination of mutivariate distributions
that we will use in the sequel.

\subsection{Transformation of an arbitrary random variable into a Gaussian
variable}
\begin{figure}
\begin{center}
\includegraphics[width=15cm]{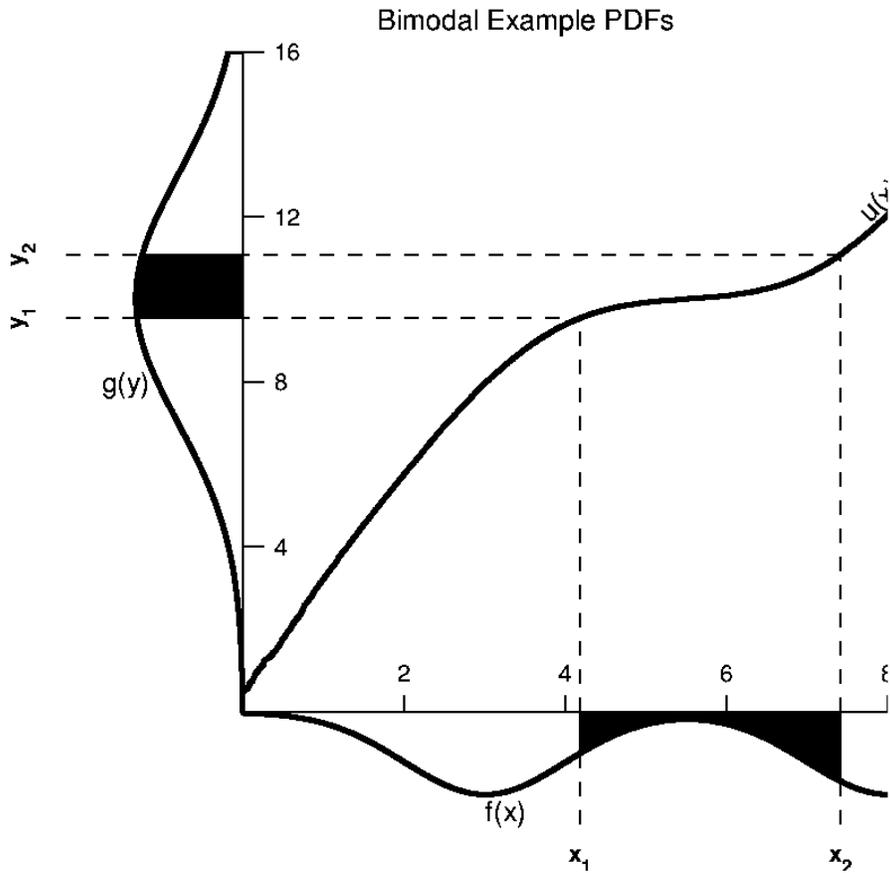}
\end{center}
\caption{\label{fig1} Schematic representation of the nonlinear
mapping $Y = u(X)$
that allows one to transform a variable $X$ with an arbitrary distribution
into a variable $Y$ with a Gaussian distribution.
The probability densities for $X$ and $Y$ are plotted
outside their respective axes. Consistent with the conservation of
probability, the shaded regions have equal area. This conservation of
probability
determines the nonlinear mapping.}
\end{figure}

Let us consider the return $X$, taken as a random variable
characterized by the probability density $p(x)$.
The transformation $y(x)$ which obtains a standard normal variable $y$ from
$x$ is
determined by the conservation of probability:
\be
p(x) dx =\frac{1}{\sqrt{2 \pi}} e^{-\frac{y^2}{2}} dy\; .
\ee
Integrating this equation from $- \infty$ and $x$, we obtain:
\be
F(x)=\frac{1}{2} \left[ 1+  \mbox{erf} \left( \frac{y}{\sqrt{2}} \right)
\right] \; ,
\ee
where $F(x)$ is the cumulative distribution of $X$:
\be
F(x) = \int_{-\infty}^x dx' p(x') \; .
\ee
This leads to the following transformation $y(x)$:
\be
\label{eq7}
y=\sqrt{2}\; \mbox{erf}^{-1}(2F(x)-1)~,
\ee
which is obvously an increasing function of $X$ as required for the
application of the
invariance property of the copula stated in the previous section.
An illustration of the nonlinear transformation (\ref{eq7}) is shown in figure
\ref{fig1}.  Note that it does not require any special hypothesis on the
probability density $X$, apart from being non-degenerate.

In the case where the pdf of $X$ has only one maximum, we may use a
simpler expression
equivalent to (\ref{eq7}). Such a pdf can be written under the
so-called Von Mises parametrization \cite{Embrecht97}~:
\be
\label{eq1}
p(x) = C \frac{f'(x)}{\sqrt{|f(x)|}} e^{-\frac{1}{2}f(x)} \; ,
\ee
where $C$ is a constant of normalization. For $f(x)/x^2 \rightarrow  0$
when $|x| \rightarrow +\infty$, the pdf has a ``fat tail,'' i.e., it
decays slower than a Gaussian at large $|x|$.

Let us now define the change of variable
\be
\label{eq2}
y=sgn(x) \sqrt{|f(x)|} \;.
\ee
Using the relationship $p(y)=p(x) \frac{dx}{dy}$, we get:
\be
\label{eq4}
p(y)=\frac{1}{\sqrt{2 \pi}} e^{-\frac{y^2}{2}} \; .
\ee
It is important to stress the presence of the sign function $sgn(x)$
in equation (\ref{eq2}), which is essential in order to
correctly quantify dependences between random variables.
This transformation (\ref{eq2}) is equivalent to (\ref{eq7})
but of a simpler implementation and will be used in the sequel.

\subsection{Determination of the joint distribution~: maximum entropy
and Gaussian copula}

Let us now consider $N$ random variables $X_i$ with marginal distributions
$p_i(x_i)$. Using the transformation
(\ref{eq7}), we define $N$ standard normal variables $Y_i$. If these variables
were independent, their joint distribution would simply be the product of
the marginal distributions. In many situations, the variables are not
independent and it is necessary to study their dependence.

The simplest approach is to construct their covariance matrix. Applied to
the variables $Y_i$, we are certain that the covariance matrix exists and
is well-defined since their marginal distributions are Gaussian.
In contrast, this is not ensured for the variables $X_i$.
Indeed, in many situations in nature, in economy, finance and in social
sciences,
pdf's are found to have power law tails  $\sim \frac{A}{x^{1+\mu}}$
for large $|x|$. If  $\mu  \le 2$, the variance and the covariances can not be
defined. If  $2 < \mu  \le 4$, the variance and the covariances exit
in principle
but their sample estimators converge poorly.

We thus define the covariance matrix:
\be
V=E[{\bf y}{\bf  y^t}] \; ,
\ee
where ${\bf y}$ is the vector of variables $Y_i$ and the operator $E[\cdot]$
represents the mathematical expectation.
A classical result of information theory \cite{Rao} tells us that,
given the covariance matrix $V$, the best joint
distribution (in the sense of entropy maximization) of the
$N$ variables $Y_i$ is the multivariate Gaussian:
\be
P({\bf y}) = \frac{1}{(2 \pi)^{N/2} \sqrt{\det(V)}} \exp \left(-\frac{1}{2}
{\bf y^t}V^{-1}{\bf y}
\right) \; .
\ee
Indeed, this distribution implies the minimum additional information or
assumption, given the covariance matrix.

Using the joint distribution of the variables $Y_i$, we obtain the joint
distribution of the variables $X_i$:
\be
P({\bf x}) = P({\bf y}) \left| \frac{ \partial y_i}{\partial x_j} \right| \; .
\ee
Since
\be
\frac{ \partial y_i}{\partial x_j} =\sqrt{2\pi} p_j(x_j) e^{\frac{1}{2} y_i^2}
\delta_{ij} \; , \ee
we get
\be
\left| \frac{ \partial y_i}{\partial x_j} \right| = (2\pi)^{N/2} \prod_{i=1}^N
p_i(x_i) e^{\frac{1}{2}  y_i^2} \; .
\ee
This finally yields
\be
P({\bf x}) = \frac{1}{\sqrt{\det(V)}} \exp \left(-\frac{1}{2} {\bf
y_{(x)}^t}(V^{-1}-I){\bf y_{(x)}}\ \right)\prod_{i=1}^N p_i(x_i) \; .
\label{jfjma}
\ee
As expected, if the variables are independent, $V=I$, and $P({\bf x})$
becomes the product of the marginal distributions of the variables $X_i$.

Let $F({\bf x})$ denote the cumulative distribution function of {\bf x} and
$F_i(x_i)$ their marginal distributions. The copula $C$ is then such that
\be
F(x_1, \cdots, x_n) = C(F_1(x_1), \cdots, F_n(x_n))~.
\ee

Differentiating with respect to $x_1, \cdots, x_N$ leads to
\be
P(x_1, \cdots, x_n) = \frac{\partial F(x_1, \cdots, x_n)}{\partial x_1 \cdots
\partial x_n} = c(F_1(x_1), \cdots, F_n(x_n))\prod_{i=1}^N p_i(x_i)~,
\ee
where
\be
c(u_1,\cdots, u_N) = \frac{\partial C(u_1,\cdots, u_N)}{\partial u_1 \cdots
\partial u_N}  \label{vfjnqloaq}
\ee
is the density of the copula $C$.

Comparing (\ref{vfjnqloaq}) with (\ref{jfjma}), the density of the
copula is given
in the present case by
\be
c(u_1,\cdots, u_N)= \frac{1}{\sqrt{\det(V)}} \exp \left(-\frac{1}{2} {\bf
y_{(u)}^t}(V^{-1}-I){\bf y_{(u)}}\ \right)~,
\ee
which is the ``Gaussian copula'' with covariance matrix ${\bf V}$. This result
clarifies and justifies the method of \cite{Sornette1} by showing that it
essentially amounts to assume arbitrary marginal distributions with Gaussian
copulas. Note that the Gaussian copula results directly from the
choice of maximizing
the Shannon entropy. This is not unexpected in analogy with
the standard result that the Gaussian law is minimizing the Shannon
entropy at fixed
given variance. If we were to extend this formulation by considering
more general
expressions of the entropy, such that Tsallis entropy \cite{Tsallisfund},
we would have found other copulas.

\subsection{Empirical test of the Gaussian copula assumption}

We now present preliminary tests of the hypothesis of Gaussian copulas
between returns of financial assets.
Testing the gaussian copula hypothesis is a delicate task. {\it A priori}, two
standard methods can be proposed.

\begin{figure}
\begin{center}
\includegraphics[width=15cm]{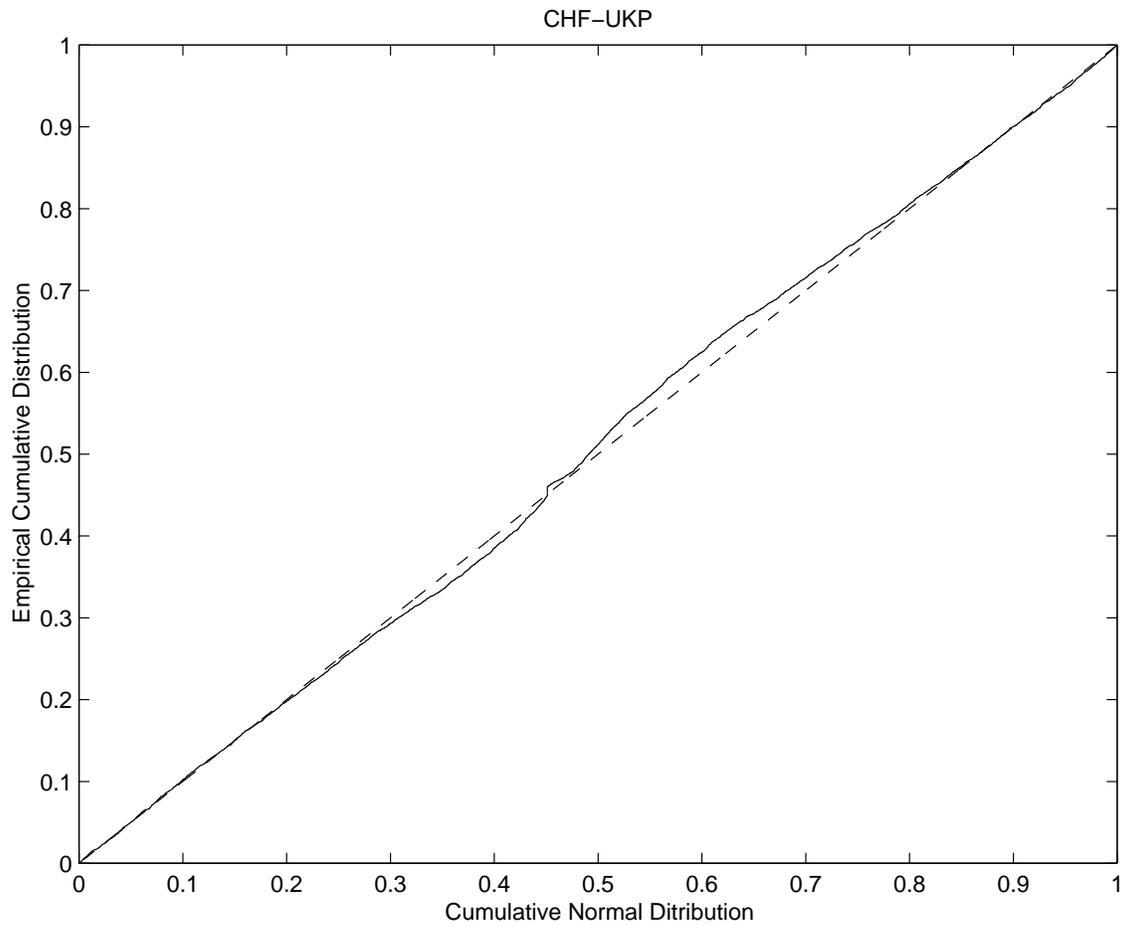}
\end{center}
\caption{\label{fig:sum_chf_ukp} Quantile of the normalized sum of the
Gaussianized returns of the Swiss Franc and The British Pound versus
the quantile of the Normal distribution, for the time interval from Jan. 1971
to Oct. 1998. Different weights in the sum give similar results.} \end{figure}

\begin{figure}
\begin{center}
\includegraphics[width=15cm]{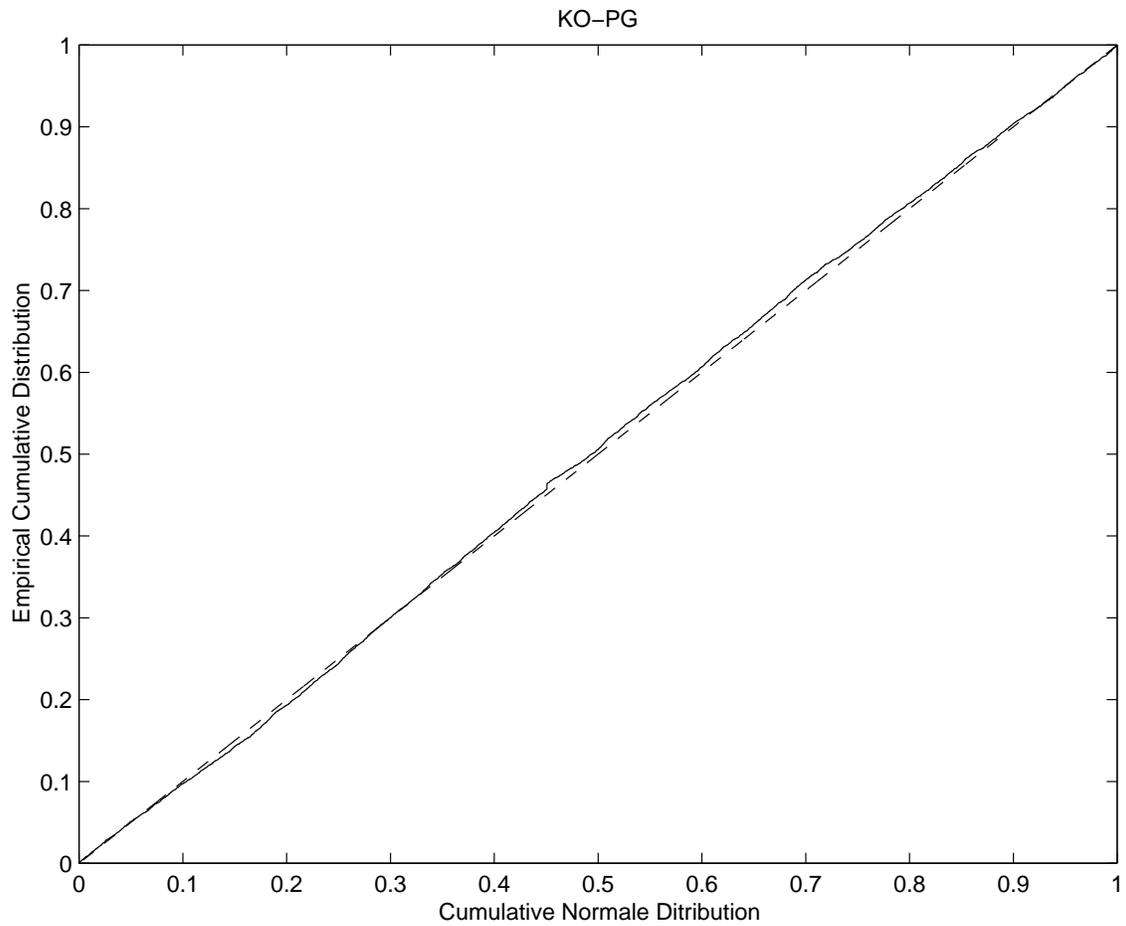}
\end{center}
\caption{\label{fig:sum_ko_pg} Quantile of the normalized sum of the
Gaussianized returns of Coca-Cola and Procter\&Gamble  versus
the quantile of the Normal distribution, for the time interval from Jan. 1970
to Dec. 2000. Different weights in the sum give similar results.}
\end{figure}

\begin{figure}
\begin{center}
\includegraphics[width=15cm]{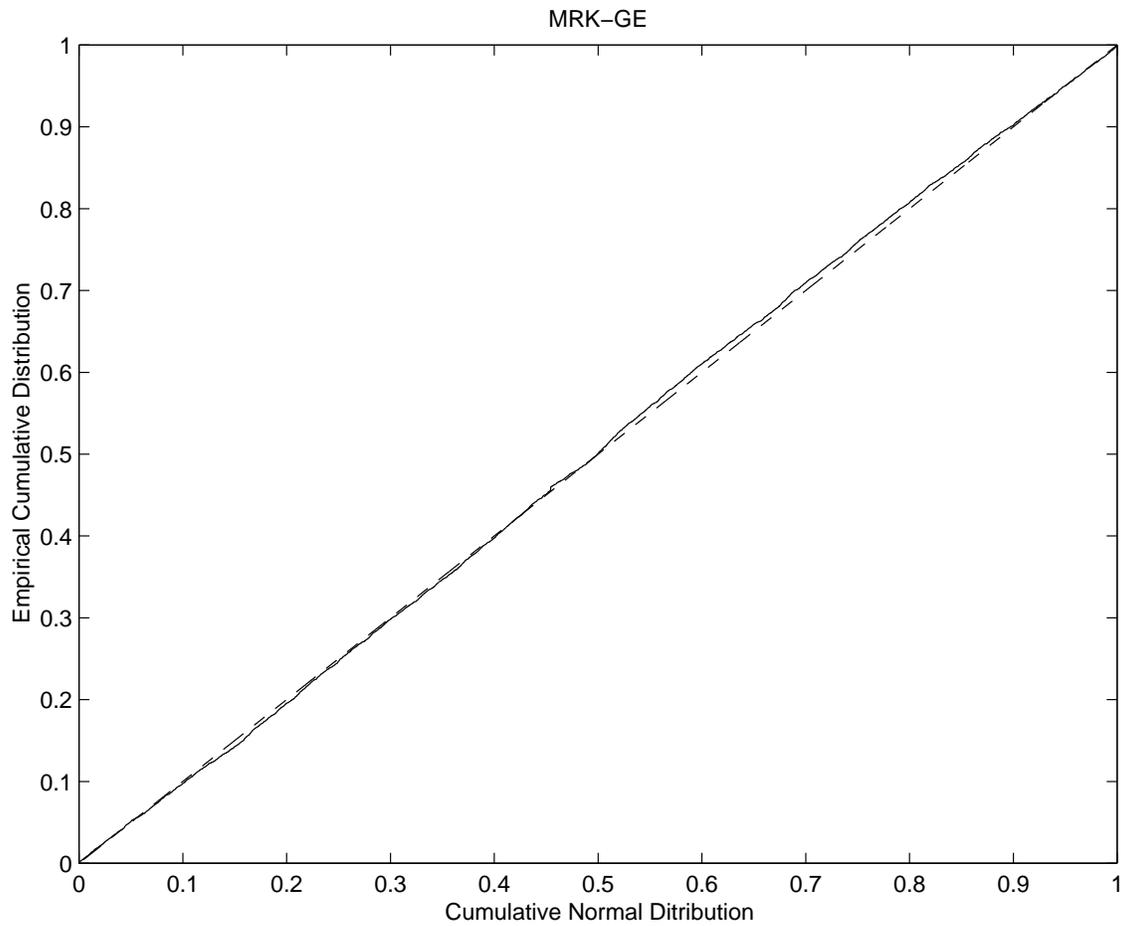}
\end{center}
\caption{\label{fig:sum_mrk_ge} Quantile of the normalized sum of the
Gaussianized returns of Merk and General Electric  versus
the quantile of the Normal distribution, for the time interval from Jan. 1970
to Dec. 2000. Different weights in the sum give similar results.}
\end{figure}

The first one consists in using that Gaussian variables are stable in
distribution under
addition. Thus, a (quantile-quantile or $Q-Q$) plot of the cumulative
distribution of the sum
$y_1+\cdots+y_p$  versus the cumulative Normal distribution with the same
estimated variance should give a straight line in order to qualify a
multivariate Gaussian
distribution (for the transformed $y$ variables). Such tests on empirical data
are presented in figures \ref{fig:sum_chf_ukp}-\ref{fig:sum_mrk_ge}.

\begin{figure}
\begin{center}
\includegraphics[width=15cm]{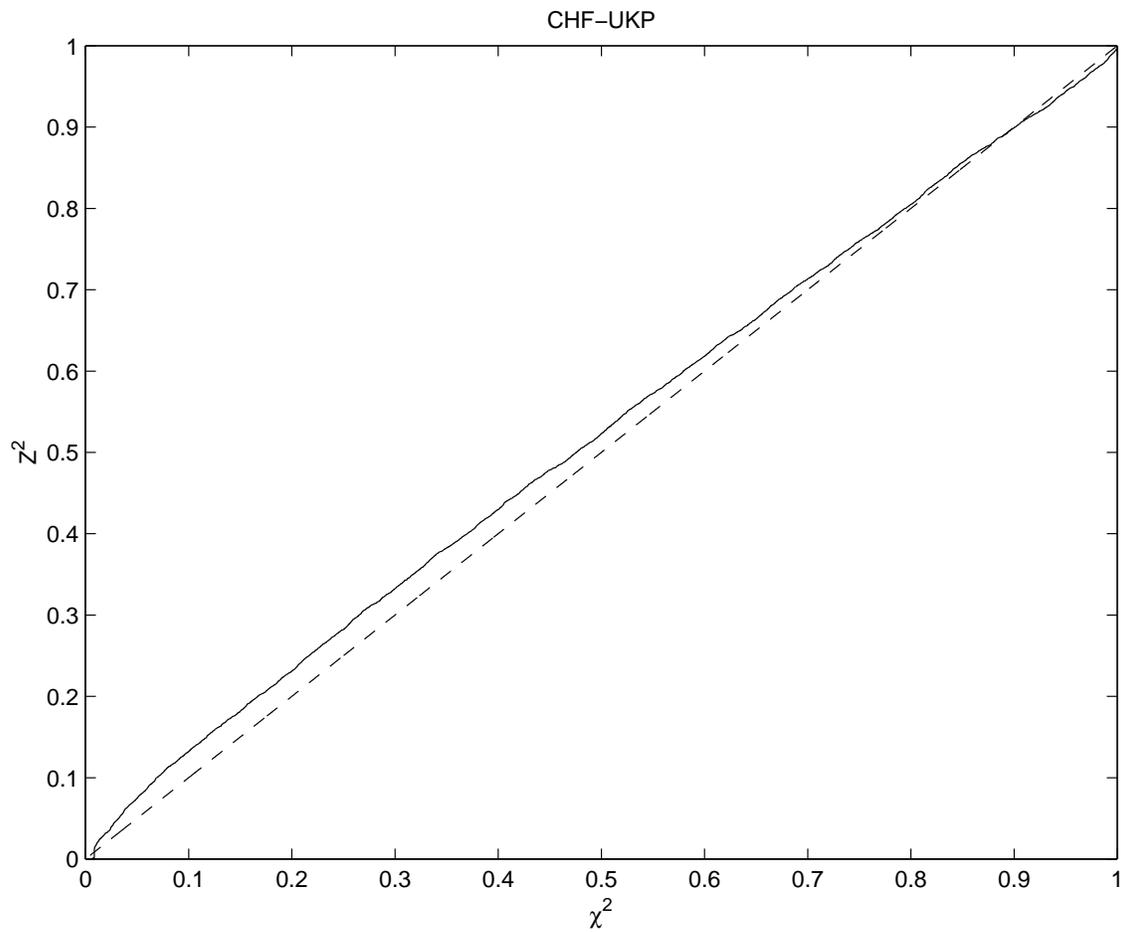}
\end{center}
\caption{\label{fig:chi2_chf_ukp} Cumulative distribution of
$ z^2={\bf y^t V^{-1} y }$ versus the cumulative distribution of
chi-square (denoted
$\chi^2$) with two degrees of freedom for the couple Swiss Franc / British
Pound, for the time interval from Jan. 1971 to Oct. 1998. This
$\chi^2$ should not be confused with
the characteristic scale used in the definition of the modified
Weibull distributions.}
\end{figure}

\begin{figure}
\begin{center}
\includegraphics[width=15cm]{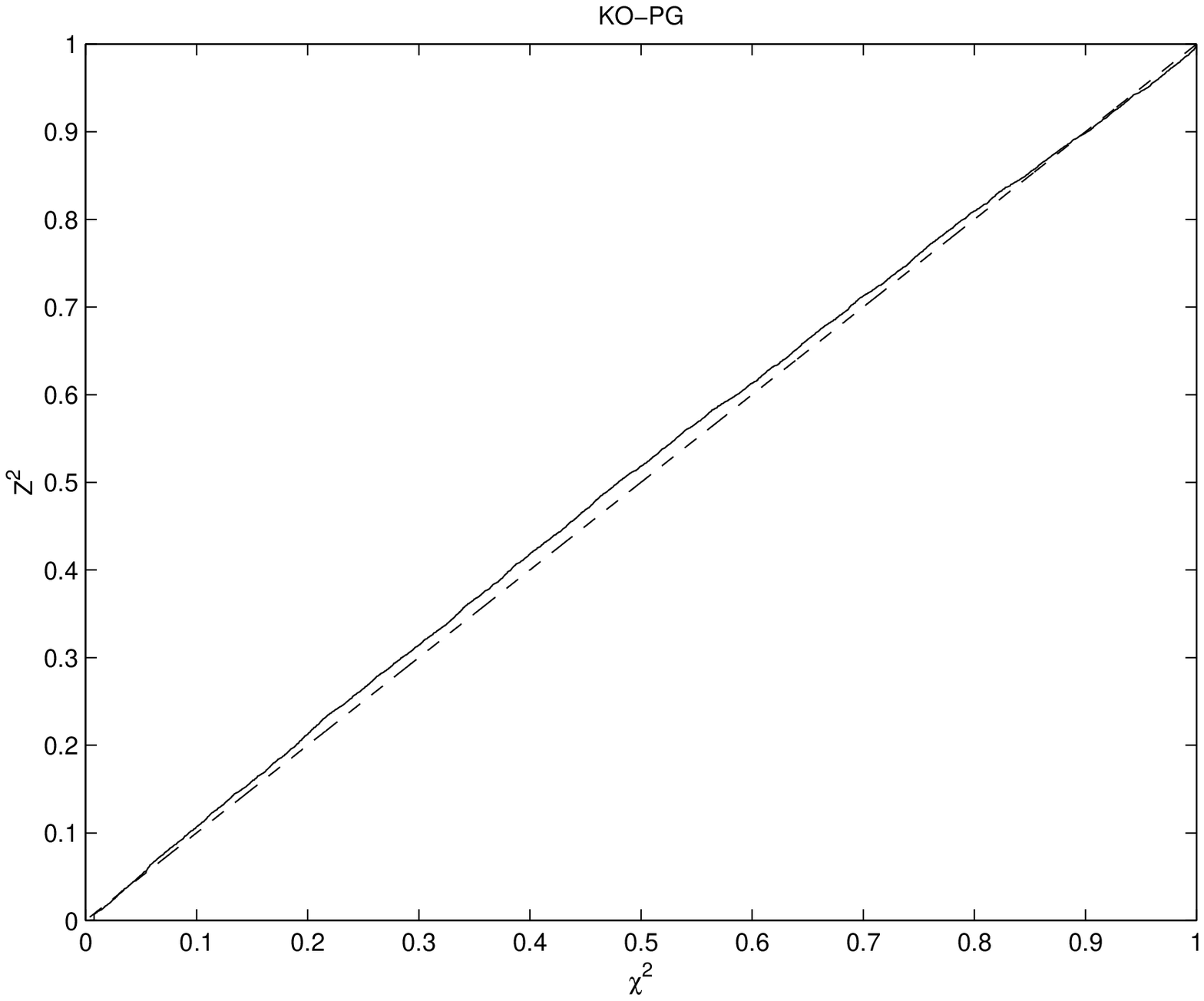}
\end{center}
\caption{\label{fig:chi2_ko_pg} Cumulative distribution of
$z^2={\bf y^t V^{-1} y}$  versus versus the cumulative distribution
of the chi-square $\chi^2$
with two degrees of freedom for the couple Coca-Cola /
Procter\&Gamble, for the time interval from Jan. 1970
to Dec. 2000. This $\chi^2$ should not be confused with
the characteristic scale used in the definition of the modified
Weibull distributions.}
\end{figure}

\begin{figure}
\begin{center}
\includegraphics[width=15cm]{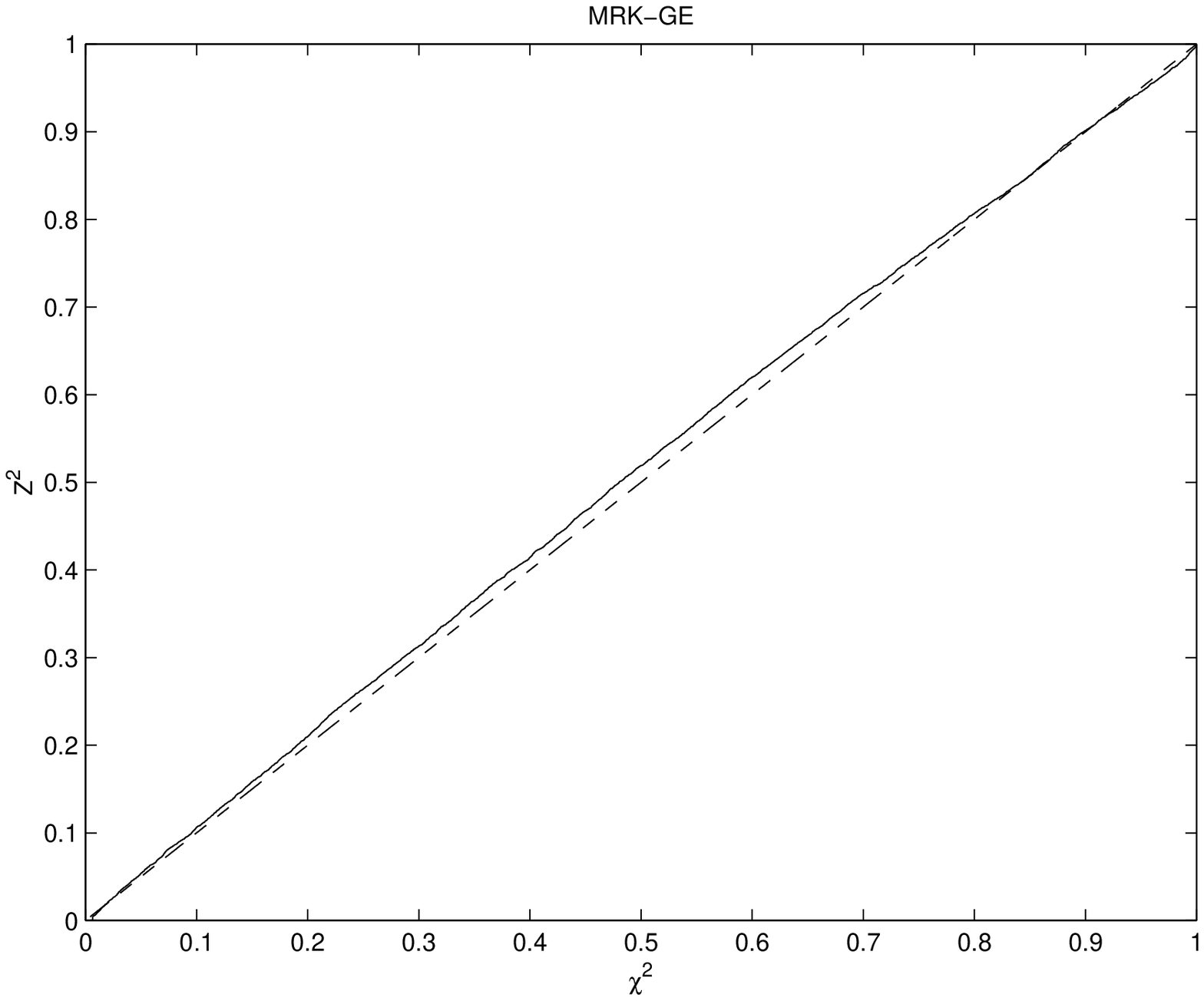}
\end{center}
\caption{\label{fig:chi2_mrk_ge} Cumulative distribution of $z^2={\bf
y^t V^{-1} y}$ versus versus the cumulative distribution of the
chi-square $\chi^2$
with two degrees of freedom for the couple Merk / General Electric,  for the
time interval from Jan. 1970 to Dec. 2000. This $\chi^2$ should not
be confused with
the characteristic scale used in the definition of the modified
Weibull distributions.}
\end{figure}

The second test amounts to estimating the covariance matrix ${\bf V}$  of the
sample we consider. This step is simple since, for fast decaying pdf's, robust
estimators of the covariance matrix are available. We can then estimate the
distribution of the variable $z^2={\bf y^t V^{-1} y}$. It is well
known that $z^2$ follows a $\chi^2$ distribution if ${\bf y}$
is a Gaussian random vector. Again, the empirical cumulative
distribution of $z^2$ versus the cumulative distribution of $\chi^2$
should give a straight line in order to qualify a multivariate Gaussian
distribution (for the transformed $y$ variables). Such tests on empirical data
are presented in figures \ref{fig:chi2_chf_ukp}-\ref{fig:chi2_mrk_ge}.

First, one can observe that the Gaussian copula
hypothesis appears better for stocks than for currencies. Note also that
the test of aggregation seems systematically more in favor of the Gaussian
copula hypothesis than is the $\chi^2$
test, maybe due to its smaller sensitivity. Nonetheless, the very good
performance of the Gaussian hypothesis under
the aggregation test bears good news for a porfolio theory based on it,
since by definition a portfolio corresponds to asset aggregation.
Even if sums of the transformed returns are not equivalent to sums
of returns (as we shall see in the sequel), such sums qualify the collective
behavior whose properties are controlled by the copula.

Notwithstanding some deviations from linearity in
figures \ref{fig:sum_chf_ukp}-\ref{fig:chi2_mrk_ge}, it appears
that, for our purpose of developing a generalized portfolio theory,
the Gaussian
copula hypothesis is a good approximation.
A more systematic test of this goodness of fit requires the quantification of a
confidence level, for instance using the Kolmogorov test,
that would allow us to accept or reject the Gaussian copula hypothesis.
However, we have encountered a practical problem to implement this test, as
the same sample is used both for the empirical calibration of the covariance
and for the test itself. A bootstrap method is thus necessary in order to
obtain a real
measure of the departure from Gaussianity and thus to decide wether or
not these deviations are significant.

\section{Choice of an exponential family to parameterize the marginal
distributions}
\subsection{The modified Weibull distributions}

We now apply these constructions to a class of distributions with fat tails,
that have been found to provide a convenient and flexible parameterization of
many phenomena found in nature and in the social sciences
\cite{Laherrere}. These so-called stretched exponential distributions
can be seen to be general forms of the extreme tails of product of
random variables
\cite{FS97}.

Following \cite{Sornette1}, we postulate the following marginal
probability distributions
of returns:
\be
\label{eq:weibull}
p(x)=\frac{1}{2\sqrt{\pi}} \frac{c}{\chi^{\frac{c}{2}}}
|x|^{\frac{c}{2}-1} e^{-\left( \frac{|x|}{\chi}\right)^c} \; ,
\ee
where $c$ and $\chi$ are the two key parameters. A more general
parameterization taking
into account a possible asymmetry between negative and positive
returns (thus leading to possible non-zero average return) is
\bea
\label{eq:pasym1}
p(x)&=&\frac{1}{2\sqrt{\pi}} \frac{c_+}{\chi_+^{\frac{c_+}{2}}}
|x|^{\frac{c_+}{2}-1} e^{-\left(
\frac{|x|}{\chi_+}\right)^{c_+}}~~~~\mbox{if}~x\ge 0\\
\label{eq:pasym2}
p(x)&=&\frac{1}{2\sqrt{\pi}} \frac{c_-}{\chi_-^{\frac{c_-}{2}}}
|x|^{\frac{c_-}{2}-1} e^{-\left(
\frac{|x|}{\chi_-}\right)^{c_-}}~~~~\mbox{if}~x<0~.
\eea
Thes expressions are close to the
Weibull distribution, with the addition of a power law prefactor to the
exponential such that the Gaussian law is retrieved for $c=2$.
Following \cite{Sornette1,SorAnderSim2,AndersenRisk},
we call (\ref{eq:weibull}) the modified Weibull distribution.
For $c<1$, the pdf is a stretched exponential, also called sub-exponential.
The exponent $c$ determines the shape of the
distribution, fatter than an exponential if $c<1$. The parameter $\chi$
controls the scale or characteristic width of the distribution. It plays
a role analogous to the standard deviation of the Gaussian law.
See chapter 6 of \cite{S2000} for a recent review on maximum likelihood
and other estimators of such generalized Weibull distributions.

\subsection{Transformation of the modified Weibull pdf into a Gaussian Law}

One advantage of the class of distributions (\ref{eq:weibull}) is that the
transformation into a Gaussian is particularly simple. Indeed, the expression
(\ref{eq:weibull}) is of the form (\ref{eq1}) with
\be
f(x)=2 \left( \frac{|x|}{\chi} \right) ^c~.
\ee
Applying the change of variable (\ref{eq2}) which reads
\be
 \label{eq:yx}
y_i=sgn(x_i)~\sqrt{2}~\left(\frac{ |x_i|}{\chi_i}\right)^{\frac{c_i}{2}} \;,
\ee
leads automatically to a Gaussian distribution.

These variables $Y_i$ then allow us to obtain the covariance matrix $V$ :
\be
V_{ij}=\frac{2}{T} \sum_{n=1}^T sgn(x_ix_j) \left( \frac{|x_i|}{\chi_i}
\right)^{\frac{c_i}{2}} \left( \frac{|x_j|}{\chi_j} \right)^{\frac{c_j}{2}}
\; ,
\ee
and thus the multivariate distributions $P({\bf y})$ and $P({\bf x})$ :
\be
\label{eq:px}
P(x_1, \cdots , x_N) = \frac{1}{2^N \pi^{N/2}\sqrt{V}} \prod_{i=1}^N
\frac{c_i |x_i|^{c/2-1}}{\chi_i^{c/2}} \exp\left[-\sum_{i,j}
V_{ij}^{-1}\left(\frac{|x_i|}{\chi_i}\right)^{c/2}
\left(\frac{|x_j|}{\chi_j}\right)^{c/2} \right]~.
\ee
Similar transforms hold, {\it mutatis mutandis}, for the asymmetric case.

\subsection{Empirical tests and estimated parameters}

\begin{figure}
\begin{center}
\includegraphics[width=15cm]{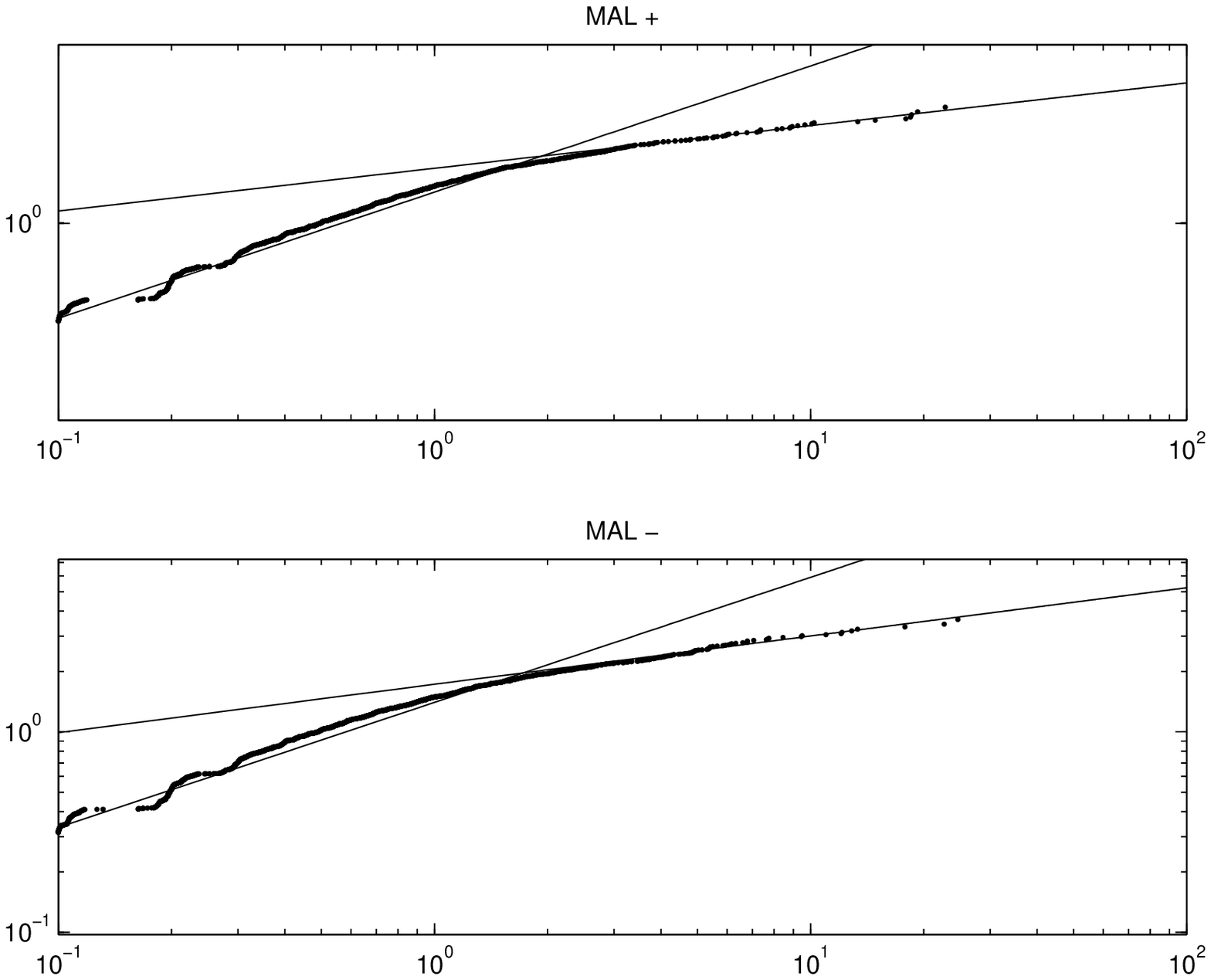}
\end{center}
\caption{\label{figYMAL} Graph of Gaussianized Malaysian Ringgit returns versus
Malaysian Ringgit returns, for the time interval from Jan. 1971
to Oct. 1998. The upper graph gives the positive tail and the
lower one the negative tail. The two straight lines represent the curves
$y=\sqrt{2} \left( \frac{x}{\langle \chi_\pm \rangle} \right)^{\langle c_\pm
\rangle}$ and $y=\sqrt{2} \left( \frac{x}{\chi_\pm} \right)^{c_\pm}$}
\end{figure}

\begin{figure}
\begin{center}
\includegraphics[width=15cm]{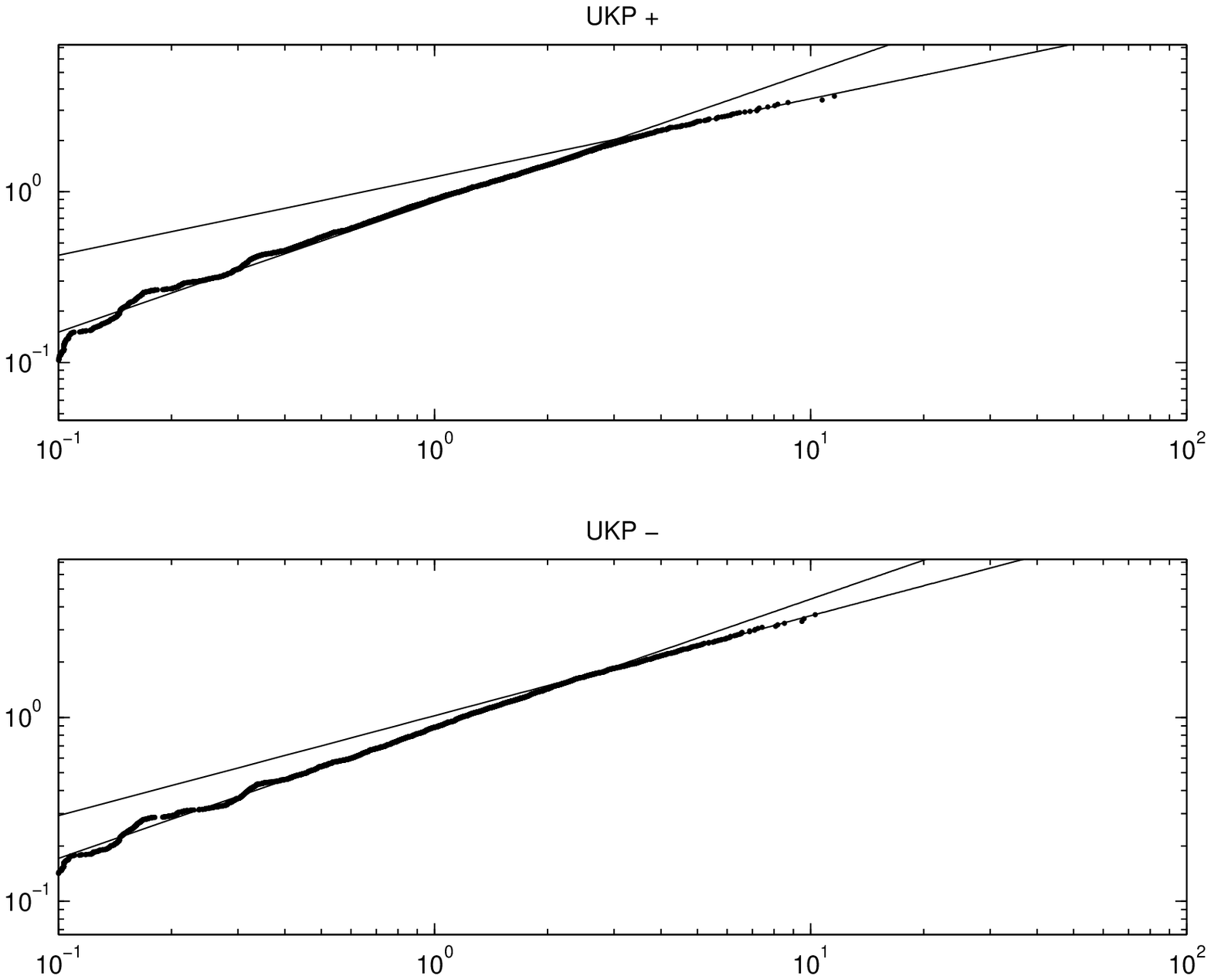}
\end{center}
\caption{\label{figYUKP} Graph of Gaussianized British Pound returns versus
British Pound returns, for the time interval from Jan. 1971
to Oct. 1998. The upper graph gives the positive tail and the lower
one the negative tail. The two straight lines represent the curves $y=\sqrt{2}
\left( \frac{x}{\langle \chi_\pm \rangle} \right)^{\langle c_\pm \rangle}$ and
$y=\sqrt{2} \left( \frac{x}{\chi_\pm} \right)^{c_\pm}$}
\end{figure}

\begin{figure}
\begin{center}
\includegraphics[width=15cm]{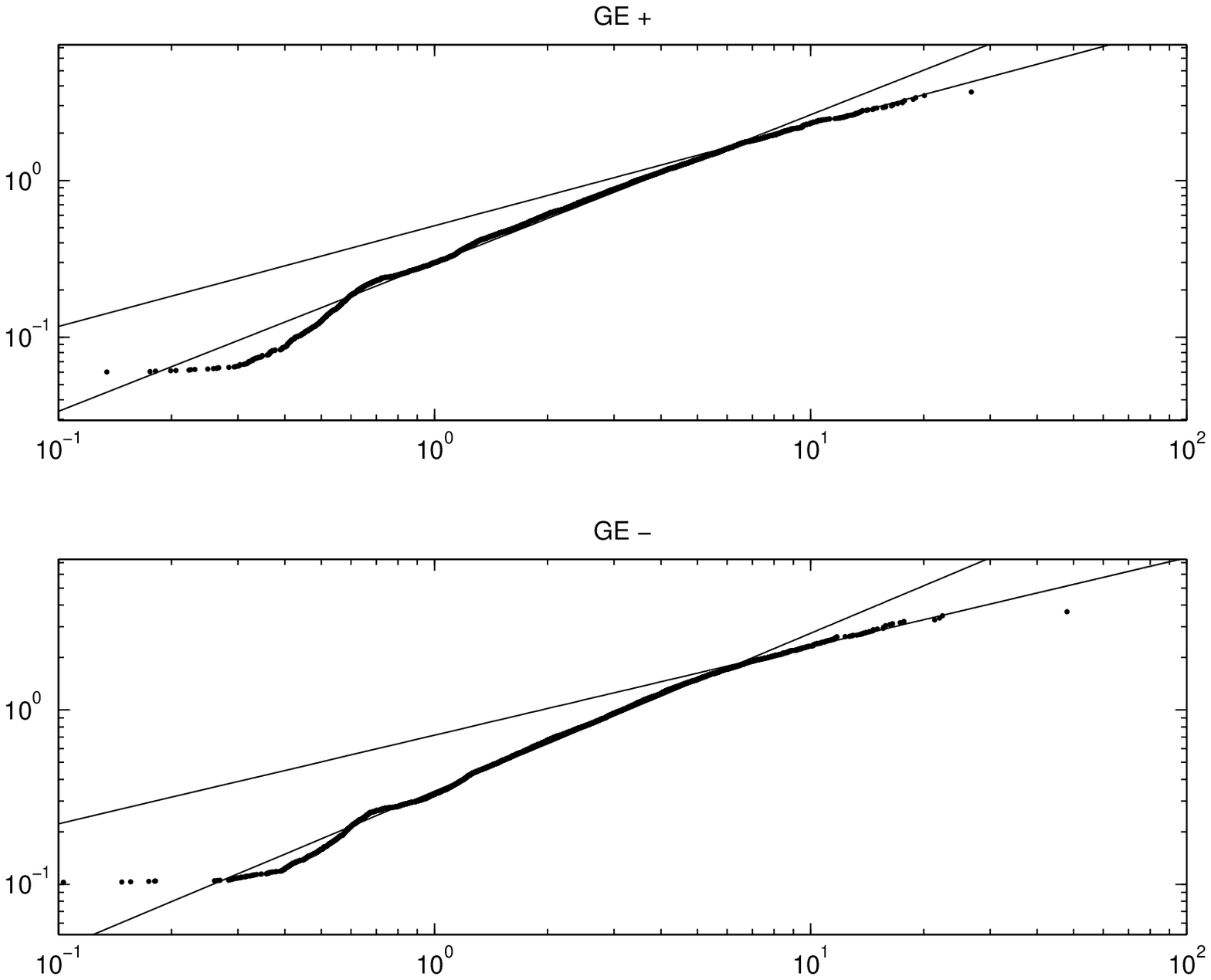}
\end{center}
\caption{\label{figYGE} Graph of Gaussianized General Electric returns versus
General Electric returns, for the time interval from Jan. 1970
to Dec. 2000. The upper graph gives the positive tail and the lower
one the negative tail. The two straight lines represent the curves $y=\sqrt{2}
\left( \frac{x}{\langle \chi_\pm \rangle} \right)^{\langle c_\pm \rangle}$ and
$y=\sqrt{2} \left( \frac{x}{\chi_\pm} \right)^{c_\pm}$}
\end{figure}

\begin{figure}
\begin{center}
\includegraphics[width=15cm]{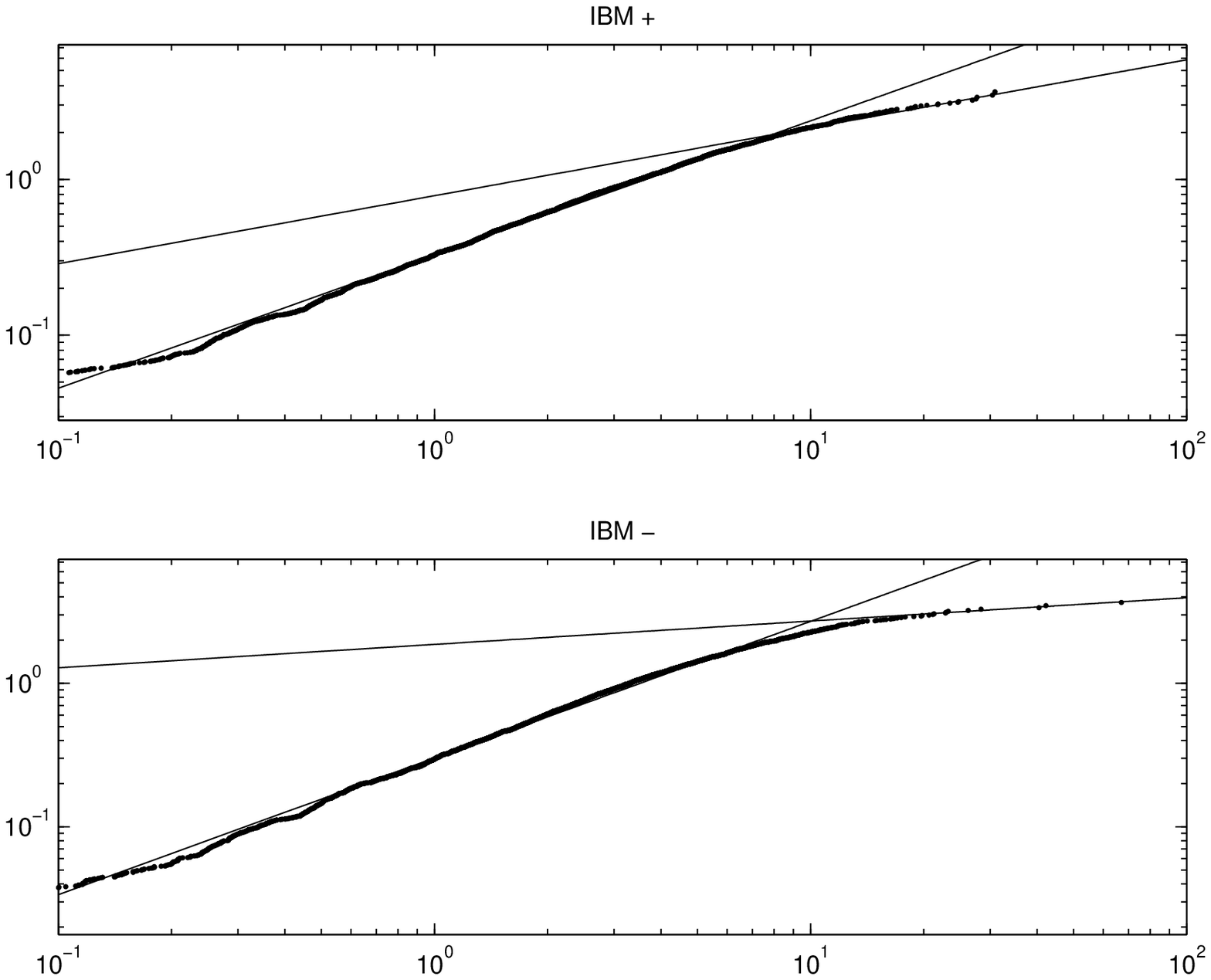}
\end{center}
\caption{\label{figYIBM} Graph of Gaussianized IBM returns versus
IBM returns, for the time interval from Jan. 1970
to Dec. 2000. The upper graph gives the positive tail and the lower
one the negative tail. The two straight lines represent the curves $y=\sqrt{2}
\left( \frac{x}{\langle \chi_\pm \rangle} \right)^{\langle c_\pm \rangle}$ and
$y=\sqrt{2} \left( \frac{x}{\chi_\pm} \right)^{c_\pm}$}
\end{figure}

\begin{figure}
\begin{center}
\includegraphics[width=15cm]{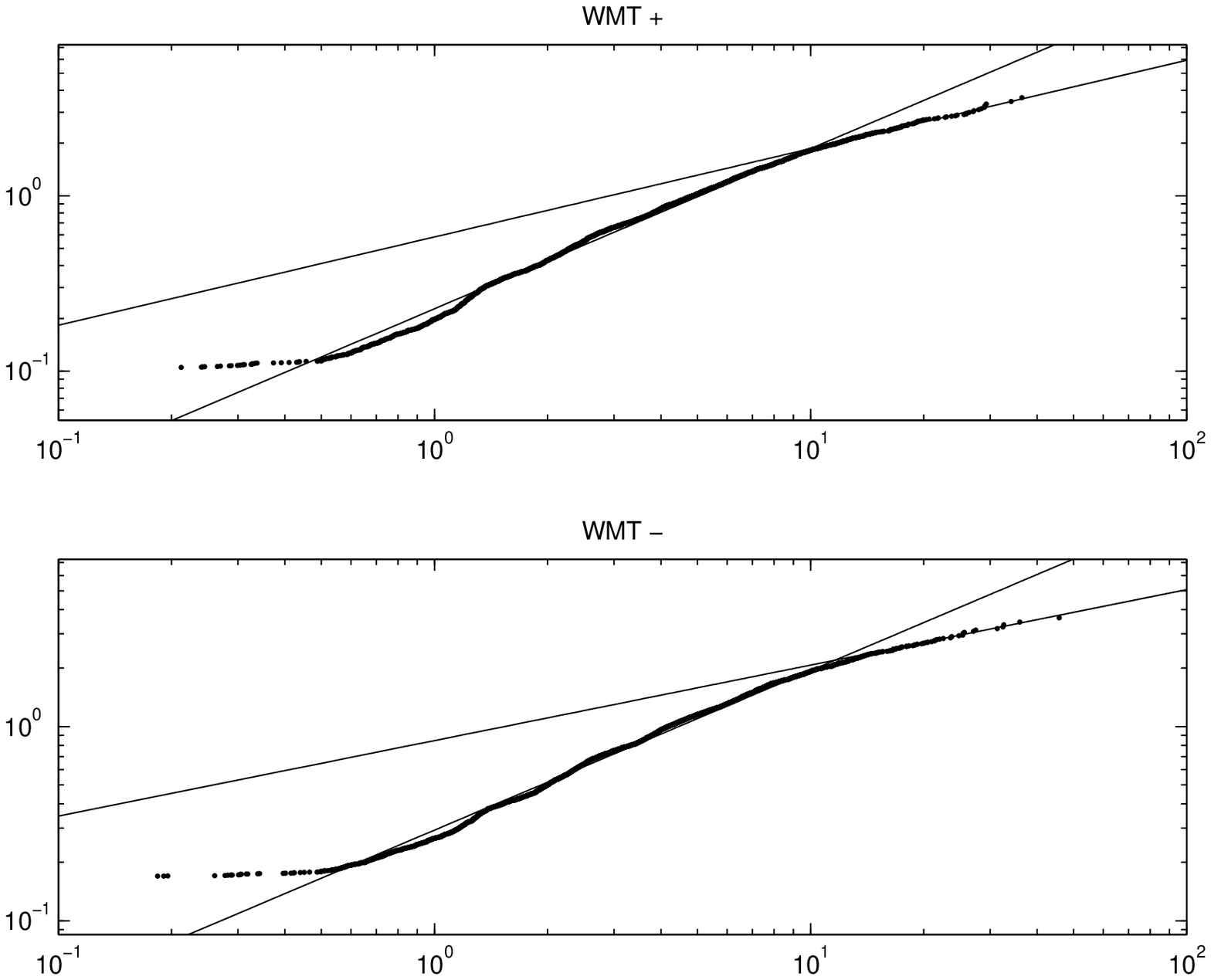}
\end{center}
\caption{\label{figYWMT} Graph of Gaussianized Wall Mart returns versus
Wall Mart returns, for the time interval from Sep. 1972
to Dec. 2000. The upper graph gives the positive tail and the lower
one the negative tail. The two straight lines represent the curves $y=\sqrt{2}
\left( \frac{x}{\langle \chi_\pm \rangle} \right)^{\langle c_\pm \rangle}$ and
$y=\sqrt{2} \left( \frac{x}{\chi_\pm} \right)^{c_\pm}$}
\end{figure}

In order to test the validity of our assumption, we have studied a
large basket of
financial assets including currencies and stocks. As an example, we present
in figures \ref{figYMAL} to \ref{figYWMT}  typical
log-log  plot of the transformed return variable $Y$ versus the
return variable $X$
for a certain number of assets. If our
assumption was right, we should observe a single straight line whose
slope is given by
$c/2$.  In contrast, we observe in general two approximately linear
regimes separated by
a cross-over. This means that the marginal distribution of returns
can be approximated
by two modified Weibull distributions, one for small returns which is
close to a Gaussian
law and one for large returns with a fat tail. Each regime is
depicted by its corresponding
straight line in the graphs. The exponents $c$ and the scale factors
$\chi$ for the different assets we have studied are given in tables
\ref{tab:1} for currencies and \ref{tab:2} for stocks. The
coefficients within brackets are the
coefficients estimated for small returns while the non-bracketed coefficients
correspond to the second fat tail regime.

The first point to note is the difference between currencies and
stocks. For small as
well as for large returns,
the exponents $c_-$ and $c_+$ for
currencies (excepted Poland and Thailand) are all close to each other.
Additional tests are required to establish whether their relatively
small differences
are statistically significant. Similarly, the scale factors are also
comparable.
In contrast, many stocks exhibit a large asymmetric behavior for large returns
with $c_+ - c_- \gtrsim 0.5$ in about one-half of the investigated stocks.
This means that the tails of the large negative returns (``crashes'')
are often much fatter
than those of the large positive returns (``rallies'').

The second important point is that, for small returns, many stocks
have an exponent
$\langle c_+ \rangle \approx \langle c_- \rangle \simeq 2$ and thus
have a  behavior not far from a
pure Gaussian, while the average exponent for currencies is about
$1.5$ in the same ``small
return'' regime.
Therefore, even for small returns, currencies exhibit
a strong departure from Gaussian behavior.

In conclusion, this empirical study
shows that the modified Weibull parameterization, although not exact on
the entire range of variation of the returns $X$, remains consistent within
each of the two regimes of small versus large returns, with a sharp
transition between
them. It seems especially relevant in the tails of the return
distributions, on which we
shall focus our attention next.

\section{Asymptotic estimation of the Value-at-Risk}
\label{sec:var}

Here, we examine theoretically the tail of the distribution
of returns of portfolios constituted of assets
with distributions characterized by the modified Weibull distributions.
Two distinct situations can occur.
\begin{itemize}
\item The tail exponents $c$ of the distributions of asset returns
are different from asset to asset.
This is the most general case and at the same time the simplest. When the
assets have different exponent $c$, the asymptotic tails of the
portfolio return
distribution are dominated by the asset with the heaviest tails.
The largest risks of the portfolio are thus controlled by the single
most risky asset
characterized by the smallest exponent $c$. Such extreme risk cannot be
diversified away. The best strategy focused on minimizing the extreme risks
in such a case consists in holding only the asset with the thinnest
tail, i.e., with
the large exponent $c$.
\item All assets in the portfolio have the same tail exponent $c$.
This case is the most interesting and challenging as we shall see.
We now present general asymptotic results for this case.
\end{itemize}

\subsection{Portfolio made of assets with the same exponent $c>1$}
\subsubsection{Case of independent assets}

Consider $N$ assets caracterized by their returns $X_i$,
$i=\{1,2,\cdots,N\}$, with joint
distribution $P({\bf x})$. We have seen in the first section that the
portfolio return $S$
can be written as
\be
S=\sum_{i=1}^N w_i X_i \; ,
\ee
where the weights $w_i$ are real coefficients. The variables $\{X_i\}$ may be
subjected to several constraints.
Then, the probability density $P_S$ of the random variable  $S$ is given by
\be
\label{eq:defmom}
P(S) = \int dx_1\cdots dx_n P(x_1, \cdots , x_N) ~\delta \left( \sum w_i x_i
-S \right)~,
\ee
where $\delta$ is the Dirac function. Our purpose here is to evaluate its
asymptotic behavior in the case where
\be
P(x_1, \cdots , x_N) = P_1(x_1) \cdots P_N(x_N)~.
\ee

Using a saddle point approximation (see appendix \ref{app:A}), we show that
for large $S$:
\be
P(S) \sim \left[\frac{c^\frac{N+1}{2}}{\chi~2^\frac{N+1}{2}
(c-1)^\frac{N-1}{2}}
\prod_{i=1}^N (w_i \chi_i) ^{\frac{c-1}{2}+\frac{1}{2(c-1)}}\right]~ \left(
\frac{|S|}{\chi} \right)^{c-2} e^{-\left( \frac{|S|}{\chi} \right)^c}~,
\ee
where $\chi$ is given by
\be
\label{eq:chi_ind}
\chi^{\frac{c}{c-1}} = \sum_{i=1}^N (w_i \chi_i)^{\frac{c}{c-1}}~.
\ee

Thus, for a large enough $VaR$ or equivalently for a small enough
loss probability,
\bea
\Pr \{S< -VaR \} &\simeq&   \frac{c^\frac{N+1}{2}}{\chi~2^\frac{N+1}{2}
(c-1)^\frac{N-1}{2}} \prod_{i=1}^N (w_i \chi_i)
^{\frac{c-1}{2}+\frac{1}{2(c-1)}}  \int_{-\infty}^{-{\rm VaR}} dS~ \left(
\frac{|S|}{\chi} \right)^{c-2} e^{-\left( \frac{|S|}{\chi} \right)^c}\\
&\simeq&  \frac{c^\frac{N-1}{2}}{2^\frac{N+1}{2}
(c-1)^\frac{N-1}{2}} \prod_{i=1}^N (w_i \chi_i)
^{\frac{c-1}{2}+\frac{1}{2(c-1)}} ~\Gamma \left(\frac{c-1}{c}, \left(
\frac{VaR}{\chi} \right)^c \right)~,\eea
where $\Gamma(\cdot, \cdot)$ is the incomplet Gamma function.
Using an asymptotic expansion of the incomplet gamma function, this finally
yields
\be
\Pr \{S< -VaR \} \simeq \left[\frac{c^\frac{N-1}{2}}{2^\frac{N+1}{2}
(c-1)^\frac{N-1}{2}} \prod_{i=1}^N (w_i \chi_i)
^{\frac{c-1}{2}+\frac{1}{2(c-1)}}\right] ~\frac{\exp\left[-\left(
\frac{VaR}{\chi}\right)^c \right]}{\left( \frac{VaR}{\chi}\right)}
\ee
Our result here recover the previously announced \cite{Sornette1}
asymptotic form
$e^{-\left( \frac{|S|}{\chi} \right)^c}$ for the tail of the distribution but
corrects an error in the calculation of the characteristic scale $\chi$.

\subsubsection{Case of dependent assets}

We now consider the case of dependent assets with pdf given by
equation (\ref{eq:px})
or its asymmetric expression. The calculation performed in appendix
\ref{app:A} leads to the following asymptotic expression for the density
function $P(S)$ for large portfolio returns $S$~:
\be
P(S)\propto
\left(\frac{|S|}{\chi}\right)^{2-c}~
e^{-\left(\frac{S}{\chi}\right)^c}~,
\ee
and the following asymptotic $VaR$~:
\be
\Pr \{S< -VaR \} \simeq   C^{st} ~\Gamma \left(\frac{c-1}{c}, \left(
\frac{VaR}{\chi} \right)^c \right)~,
\ee
where  $C^{st}$ is a term independent of $S$, whose expression is given
appendix \ref{app:A} and $\chi$ is such that
\be
\chi^c = \frac{1} {c\sum_{i,j} V_{ij}^{-1} \sigma_i^{c/2} \sigma_j^{c/2}}~,
\ee
and the $\{ \sigma_i\}$ are solution of
\be
\sum_{i}
V_{ik}^{-1} \sigma_i^{c/2} \sigma_k^{c/2}= w_k \chi_k \sigma_k ~\sum_{i,j}
V_{ij}^{-1} \sigma_i^{c/2} \sigma_j^{c/2} ~.
\ee

Obviously, if the assets are independent, $V_{ij}^{-1}=\delta_{ij}$,
the $\sigma_i$'s are solution of
\bea
\sigma_i^{c-1}&=&\lambda w_i \chi_i,~ ~ \forall i \in\{1, \cdots,
N\}\\
\lambda &=& \sum_i \sigma_i^c~,
\eea
$\chi$ is then simply given by equation (\ref{eq:chi_ind}).

\subsection{Portfolio made of assets with the same exponent $c<1$}

In the case of independent assets whose tails of  distributions are fatter
than an exponential, the tail of the distribution of price variations of the
portfolio is given by \cite{Sornette1} \be
\label{eq3.1}
P(S) \sim \exp \left[-\left( \frac{|S|}{\chi}\right)^c \right] \; ,
\ee
with
\be
\chi^2=\max_{i\in \{1,..N\}} \{w_1^2 \chi_1^2, \cdots, w_N^2 \chi_N^2\}~.
\ee
This single variable $\chi$ controls the asymptotic behavior of the tail of the
distribution.

Using the equation (\ref{eq3.1}), the Value-at-Risk is solution of
\be
\Pr \{ S <-VaR \} \sim \int_{-\infty}^{-{\rm VaR}} dS~e^{-\left(
\frac{|S|}{\chi}\right)^c }~ ,
\ee
i.e~:
\be
\Pr \{ S <-VaR \} \sim \Gamma \left( \frac{1}{c},{\rm VaR}^c
\right)~.
\ee

In the case of correlated assets, we have not yet found an asymptotic formula.
But, using the standard result that a sum of stretched exponentially
distributed random variables behaves, in the regime of large deviation, like
the variable whose deviation is the most extreme, we expect the
previous result to remain true.

\subsection{Summary}

In summary, the density function $P(S)$ behaves for large
$S$ like
\be
P(S) \sim e^{- \left(\frac{|S|}{\chi} \right)^c}~, \label{ghfla}
\ee
where
\bea
\chi &=& \left( \sum_i (w_i \chi_i )^\frac{c}{c-1} \right) ^\frac{c-1}{c}~ ~
\mbox{if }~ c>1~~\mbox{ and~ independent~ assets}~,  \label{cawseks}\\
\chi &=& \frac{1}{ \left(c\sum_{i,j} V_{ij}^{-1} \sigma_i^{c/2}
\sigma_j^{c/2}\right) ^\frac{1}{c}}~ ~ \mbox{if} ~ c>1~ ~\mbox{ and~
dependent~ assets}~,\\
\chi &=& \left( \max_{i\in \{1,..N\}} \{w_1^2 \chi_1^2, \cdots, w_N^2
\chi_N^2\} \right)^{1/2}~ ~ \mbox{if }~ c<1~.  \label{cawsaaeks}
\eea
Note that the expression (\ref{cawseks}) for $c>1$ retrieves the
result (\ref{cawsaaeks})
for $c<1$ by taking the limit $c \to 1^+$, showing the continuity of
the formulas.

\begin{figure}
\begin{center}
\includegraphics[width=15cm]{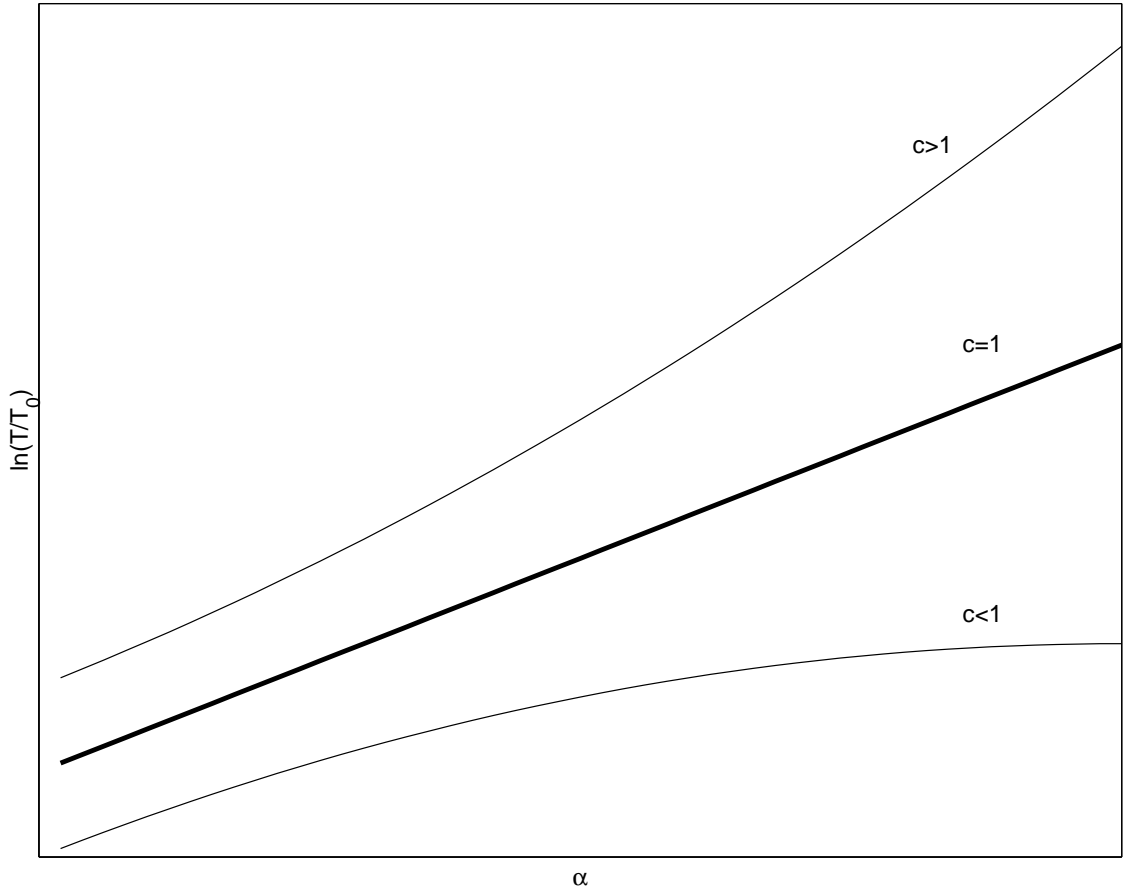}
\end{center}
\caption{\label{fig:varfreq} Logarithm $\ln \left(
\frac{T}{T_0}\right)$ of the ratio of the
recurrence time $T$ to a reference time $T_0$ for the recurrence of a
given loss $VaR$
as a function of $\alpha$ defined by $\alpha = \frac{VaR}{VaR^*}$.
$VaR^*$ (resp.
$VaR$) is the Value-at-Risk over a time interval $T_0$ (resp.
$T$). }
\end{figure}

Let us translate these formulas in intuitive form. For this, we
define a value-at-risk
(VaR) $VaR^*$ which is such that its typical frequency is $1/T_0$. $T_0$ is
by definition the typical
recurrence time of a loss larger than $VaR^*$. In our present example, we take
$T_0$ equals $1$ year for example, i.e., $VaR^*$ is the typical
annual shock or crash.
The expression (\ref{ghfla}) then allows us to predict the recurrence
time $T$ of
a VaR equal to $\alpha$ times this reference value $VaR^*$:
\be
\ln \left( \frac{T}{T_0} \right) \simeq  (\alpha^c-1) \left(
\frac{VaR^*}{\chi} \right)^c + {\cal O}(\ln \alpha)~.
\ee
Figure \ref{fig:varfreq} shows $\ln \frac{T}{T_0}$ versus $\alpha$.
For fixed $\alpha$, $T$
increases more and more slowly when the exponent $c$ decreases. This
quantifies our expectation that
large losses occur more frequently for sub-exponential distributions than
for super-exponential ones.


\section{Cumulant expansion of the portfolio return distribution}
\label{sec:cum}
\subsection{link between moments and cumulants}

Before deriving the main result of this section, we recall a standard
relation between moments and cumulants that we need below.

The moments $M_n$ of the distribution $P$ are defined by
\be
\hat P(k) = \sum_{n=0}^{+ \infty} \frac{(ik)^n}{n!}M_n~,
\ee
where $\hat P$ is the characteristic function, i.e., the Fourier transform of
$P$~:
\be
\hat P(k) = \int_{-\infty}^{+ \infty} dS~P(S) e^{ikS}~.
\ee
Similarly, the cumulants $C_n$ are given by
\be
\hat P(k) = \exp \left(\sum_{n=1}^{+ \infty} \frac{(ik)^n}{n!}C_n \right)~.
\label{ghlqjfjgf}
\ee

Differentiating $n$ times the equation
\be
\ln \left(\sum_{n=0}^{+ \infty} \frac{(ik)^n}{n!}M_n \right) = \sum_{n=1}^{+
\infty} \frac{(ik)^n}{n!}C_n \; ,
 \ee
we obtain the following recurrence relations between the moments and the
cumulants~:
\bea
M_n & = & \sum_{p=0}^{n-1}  {{n-1} \choose{p}} M_p C_{n-p} \; ,\\
\label{eq:cum}
C_n & = &  M_n - \sum_{p=1}^{n-1} {{n-1} \choose {n-p}} C_p M_{n-p} \;.
\eea

In the sequel, we will first evaluate the moments, which turns out to
be easier, and
then using eq (\ref{eq:cum}) we will be able to calculate the
cumulants. Cumulants
are indeed the natural objects to quantify risks: as seen from their definition
(\ref{ghlqjfjgf}), cumulants of order larger than $2$ quantify
deviation from the
Gaussian law, and thus large risks beyond the variance (equal to the
second-order
cumulant). More importantly, they are invariant with respect to
translations or change of
a return of reference and are thus appropriate measures of fluctuations.

\subsection{Symmetric assets}

We start with the expression (\ref{eq:defmom}) of the distribution of
the weighted
sum of $N$ assets~:
\be
P_S(s)=\int_{R^N} d{\bf x} \; P({\bf x}) \delta(\sum_{i=1}^N w_i x_i-s) \; .
\ee
Using the change of variable (\ref{eq2}), allowing us to go from the
asset returns
$X_i$'s to the transformed returns $Y_i$'s, we get
\be
P_S(s) = \frac{1}{(2 \pi)^{N/2}
\sqrt{\det(V)}} \int_{R^N} d{\bf y} \;   e^{-\frac{1}{2} {\bf
y^t}V^{-1}{\bf y} }
\; \delta(\sum_{i=1}^N w_i sgn(y_i)f^{-1}(y_i^2)-s) \; .
\ee
Taking its Fourier transform $\hat P_S(k) = \int ds P_S(s) e^{iks}$, we obtain
\be
\label{eqq}
\hat P_S(k) = \frac{1}{(2 \pi)^{N/2} \sqrt{\det(V)}}  \int_{R^N} d{\bf y}
\;   e^{-\frac{1}{2}
{\bf y^t}V^{-1}{\bf y} + ik \sum_{i=1}^N w_i sgn(y_i)f^{-1}(y_i^2)} \; ,
\ee
where $\hat P_S$ is the characteristic function of $P_S$.

In the particular case of interest here where the marginal distributions
of the variables $X_i$'s are the modified Weibull pdf,
\be
f^{-1}(y_i)=\chi_i |\frac{y_i}{\sqrt{2}}|^{q_i}
\ee
with
\be
q_i=2/c_i~,
\ee
the equation
(\ref{eqq}) becomes
\be
\label{eq2.5}
\hat P_S(k) = \frac{1}{(2 \pi)^{N/2} \sqrt{\det(V)}}  \int_{R^N} d{\bf y}
\;   e^{-\frac{1}{2}
{\bf y^t}V^{-1}{\bf y} + ik \sum_{i=1}^N w_i sgn(y_i) \chi_i
|\frac{y_i}{\sqrt{2}}|^{q_i}} \; .
\ee
The task in front of us is to evaluate this expression through the
determination
of the moments and/or cumulants.

\subsubsection{Case of independent assets}
In this case, the cumulants can be obtained explicitely \cite{Sornette1}.
Indeed, the expression (\ref{eq2.5}) can be expressed as a product of integrals
of the form
\be
\int_0^{+\infty} du \;e^{-\frac{u^2}{2}+ik w_i \chi_i \left(
\frac{u}{\sqrt{2}}\right)^{q_i}} \; .
\ee
We obtain
\be
\label{eq:sumcum}
C_{2n}=\sum_{i=1}^N c(n,q_i) (\chi_i w_i)^{2n} \; ,
\ee
and
\be
c(n,q_i) = (2n)! \left\{ \sum_{p=0}^{n-2} (-1)^n \frac{\Gamma
\left(
q_i(n-p)+\frac{1}{2} \right)}{(2n-2p)! \pi^{1/2}} \left[ \frac{\Gamma \left(
q_i+\frac{1}{2} \right)}{2! \pi^{1/2}} \right]^p - \frac{(-1)^n}{n}\left[
\frac{\Gamma \left(
q_i+\frac{1}{2} \right)}{2! \pi^{1/2}} \right] ^n \right\} \; .
\ee

Note that the coefficient $c(n,q_i)$ is the cumulant of order $n$ of the
marginal distribution (\ref{eq:weibull}) with $c=2/q_i$ and $\chi=1$. The
equation (\ref{eq:sumcum}) expresses simply the fact that the cumulants
of the sum of independent variables is the sum of the cumulants of each
variable.
The odd-order cumulants are zero due to the symmetry of the distributions.

\subsubsection{Case of dependent assets}

Here, we restrict our exposition to the case of two random variables. The
case with $N$ arbitrary can be treated in a similar way but involves
rather complex formulas. The equation (\ref{eq2.5}) reads
\bea
\hat P_S(k) = \frac{1}{2 \pi \sqrt{1-\rho^2}} \int dy_1 dy_2~\exp \left[
-\frac{1}{2} y^t V^{-1} y + ik \left( \chi_1 w_1 sgn(y_1)
\left| \frac{y_1}{\sqrt{2}}\right|^{q_1} + \right. \right. {\nonumber} \\
\left. \left. + \chi_2 w_2 sgn(y_2) \left| \frac{y_2}{\sqrt{2}}
\right|^{q_2} \right)
\right]~,
\eea
and we can show (see appendix \ref{app:B}) that the moments read
\be
\label{eq:amom}
M_n = \sum_{p=0}^n {n \choose p}  w_1^p w_2^{n-p}
\gamma_{q_1 q_2}(n,p)~,
\ee
with
\bea
\label{eq:gammasym}
\gamma_{q_1 q_2}(2n,2p) &=& \chi_1^{2p} \chi_2^{2(n-p)}\frac{\Gamma \left( q_1p
+\frac{1}{2} \right) \Gamma \left(q_2(n-p)+\frac{1}{2}
\right)}{\pi}~
{_2F_1} \left(-q_1p, -q_2(n-p) ; \frac{1}{2} ; \rho^2 \right) \; , \\
\gamma_{q_1 q_2}(2n,2p+1) &=&2\chi_1^{2p+1} \chi_2^{2(n-p)-1} \frac{\Gamma
\left( q_1p+1+\frac{q_1}{2} \right) \Gamma \left(q_2(n-p)+1-\frac{q_2}{2}
\right)}{\pi} \rho~{_2F_1}\left( -q_1p - \frac{q_1-1}{2},\right.\nonumber \\
&& \left. ,  -q_2(n-p)+ \frac{q_2+1}{2} ;
\frac{3}{2} ;\rho^2 \right) ~,
\eea
where ${_2F_1}$ is an hypergeometric function.

These two relations allow us to calculate the moments and cumulants for
any possible values of $q_1=2/c_1$ and $q_2=2/c_2$. If one of the
$q_i$'s is an integer, a
simplification occurs and the coefficients $\gamma(n,p)$ reduce to
polynomials. In the simpler case where all the $q_i$'s are odd integer the
expression of moments becomes~:
\be
\label{eq:mom2}
M_n=\sum_{p=0}^n {n \choose p} \left( w_1 {\chi_1} \right)^p \left( w_2
{\chi_2} \right)^{n-p}  \sum_{s=0}^{min\{q_1p,q_2(n-p)\}} {\rho}^s~
s!~ a_{s}^{(q_1p)} a_s^{(q_2(n-p))} ~,
\ee
with
\bea
a_{2p}^{(2n)} & = & (2p)! C_{2n}^{2(n-p)} (2(n-p)-1)!!
=\frac{(2n)!}{2^{(2n-2p)}
(2p)! (n-p)!}  \; ,\\
a_{2p+1}^{(2n)} & = & 0  \; ,\\
a_{2p}^{(2n+1)} & = & 0  \; ,\\
a_{2p+1}^{(2n+1)} & = & (2p+1)! C_{2n+1}^{2(n-p)} (2(n-p)-1)!! =
\frac{(2n+1)!}{2^{(2n-2p)} (2p+1)! (n-p)!} \; ,
\eea

\subsection{Non-symmetric assets}

In the case of asymmetric assets, we have to consider the formula
(\ref{eq:pasym1}-\ref{eq:pasym2}), and using the same notation as in the
previous section, the moments are again given by (\ref{eq:amom}) with
the coefficient $\gamma(n,p)$ now equal to~:
\bea
\gamma(n,p)=\frac{(-1)^n (\chi_1^{-})^p (\chi_2^{-})^{n-p}}{4 \pi} \left[
\Gamma\left( \frac{q_1^- p +1}{2} \right) \Gamma\left(\frac{q_2^-(n-p)+1}{2}
\right)~_2F_1 \left( - \frac{q_1^- p}{2} , - \frac{q_2^- (n-p)}{2} ;
\frac{1}{2} ; \rho^2 \right) +  \right. \nonumber \\
\left. + 2 \Gamma\left(
\frac{q_1^- p}{2} + 1 \right) \Gamma\left(\frac{q_2^-(n-p)}{2} + 1
\right)\rho ~_2F_1 \left( - \frac{q_1^- p - 1}{2} , - \frac{q_2^- (n-p) - 1}{2}
; \frac{3}{2} ; \rho^2 \right) \right] +  \nonumber \\
+ \frac{ (-1)^p(\chi_1^{-})^p (\chi_2^{+})^{n-p}}{4 \pi} \left[ \Gamma\left(
\frac{q_1^- p +1}{2} \right) \Gamma\left(\frac{q_2^+(n-p)+1}{2}
\right)~_2F_1 \left( - \frac{q_1^- p}{2} , - \frac{q_2^+ (n-p)}{2} ;
\frac{1}{2} ; \rho^2 \right) +  \right. \nonumber \\
\left . - 2 \Gamma\left(
\frac{q_1^- p}{2} + 1 \right) \Gamma\left(\frac{q_2^+(n-p)}{2} + 1
\right)\rho ~_2F_1 \left( - \frac{q_1^- p - 1}{2} , - \frac{q_2^+ (n-p) - 1}{2}
; \frac{3}{2} ; \rho^2 \right) \right] +  \nonumber \\
+ \frac{(-1)^{n-p}(\chi_1^{+})^p (\chi_2^{-})^{n-p}}{4 \pi} \left[ \Gamma\left(
\frac{q_1^+ p +1}{2} \right) \Gamma\left(\frac{q_2^-(n-p)+1}{2}
\right)~_2F_1 \left( - \frac{q_1^+ p}{2} , - \frac{q_2^- (n-p)}{2} ;
  \frac{1}{2} ; \rho^2 \right) +  \right. \nonumber \\
\left . - 2 \Gamma\left(
\frac{q_1^+ p}{2} + 1 \right) \Gamma\left(\frac{q_2^-(n-p)}{2} + 1
\right)\rho ~_2F_1 \left( - \frac{q_1^+ p - 1}{2} , - \frac{q_2^- (n-p) -
1}{2} ;   \frac{3}{2} ; \rho^2 \right) \right]+  \nonumber \\
+\frac{(\chi_1^{+})^p (\chi_2^{+})^{n-p}}{4 \pi} \left[ \Gamma\left(
\frac{q_1^+ p +1}{2} \right) \Gamma\left(\frac{q_2^+(n-p)+1}{2}
\right)~_2F_1 \left( - \frac{q_1^+ p}{2} , - \frac{q_2^+ (n-p)}{2} ;
  \frac{1}{2} ; \rho^2 \right) +  \right. \nonumber \\
\left . + 2 \Gamma\left(
\frac{q_1^+ p}{2} + 1 \right) \Gamma\left(\frac{q_2^+(n-p)}{2} + 1
\right)\rho ~_2F_1 \left( - \frac{q_1^+ p - 1}{2} , - \frac{q_2^+ (n-p) - 1}{2}
;   \frac{3}{2} ; \rho^2 \right) \right] \nonumber~.\\
\eea
This formula is obtained in the same way as the formulas given in the
symmetric case. We retrieve the formula (\ref{eq:gammasym}) as it should if the
coefficients '+' are equal to the coefficients '-'.

\subsection{Empirical tests}

Extensive tests have been performed for currencies under the assumption that
the distributions of asset returns are symmetric \cite{Sornette1}.

As an exemple, let us consider
the Swiss franc and the Japanese Yen against the US dollar. The calibration of
the modified Weibull distribution to the tail of the empirical histogram of
daily returns give $(q_{CHF}=1.75 , c_{CHF}=1.14, \chi_{CHF}=2.13)$ and
$(q_{JPY}=2.50 , c_{JPY}=0.8, \chi_{JPY}=1.25)$ and their correlation
coefficient is
$\rho=0.43$.

\begin{figure}
\begin{center}
\includegraphics[width=15cm]{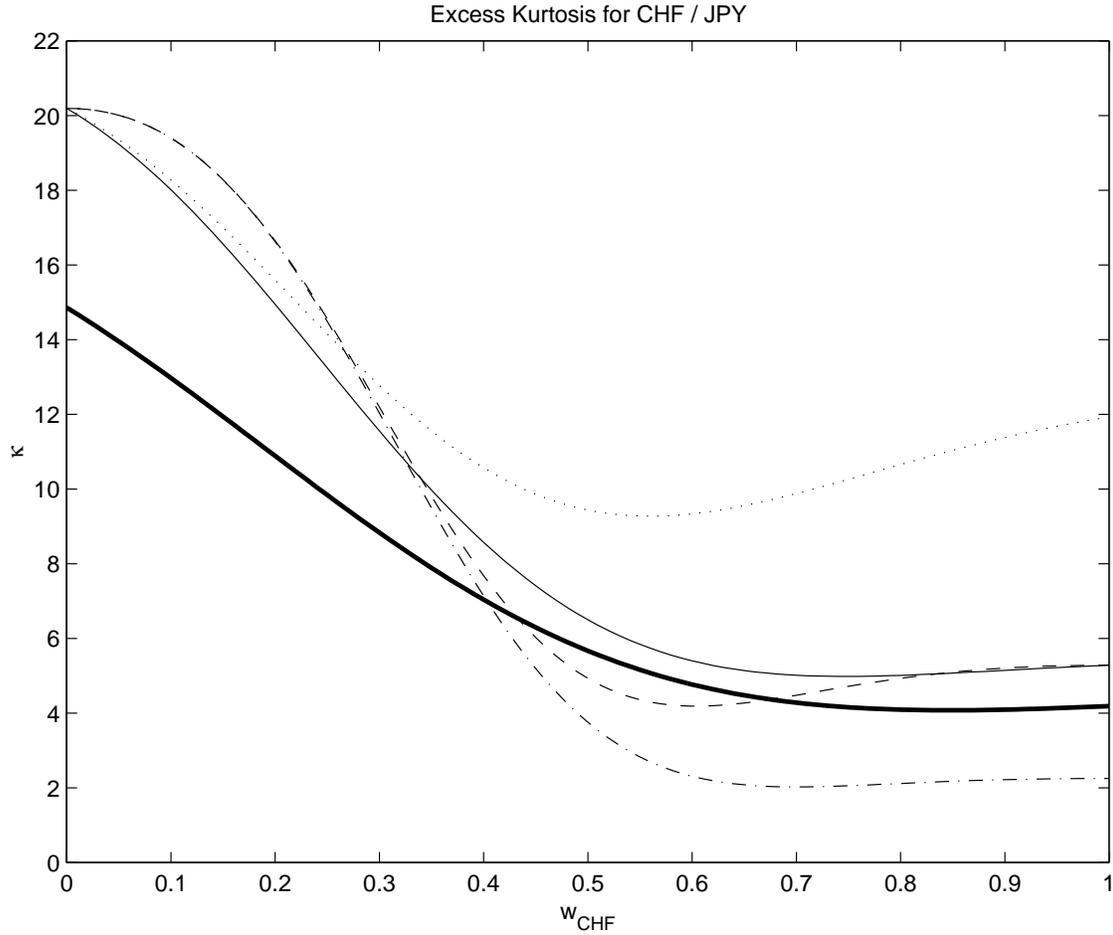}
\end{center}
\caption{\label{fig2.2} Excess kurtosis of the distribution of the price
variation $w_{CHF} x_{CHF} + w_{JPY}
x_{JPY}$ of the portfolio made of
a fraction $w_{CHF}$ of Swiss franc and a fraction
$w_{JPY}=1-w_{CHF}$ of the Japanese Yen against the US dollar, as a function of
$w_{CHF}$, denoted $w$ in the figure.
Thick solid line : empirical curve, thin solid line : theoretical curve, dashed
line : theoretical curve with $\rho=0$ (instead of $\rho=0.43$), dotted line:
theoretical curve with $q_{CHF}=2$ rather than $1.75$ and dashed-dotted line:
theoretical curve with $q_{CHF}=1.5$. The excess kurtosis has been evaluated
for the time interval from Jan. 1971 to Oct. 1998.}
\end{figure}

The figure \ref{fig2.2} plots the excess kurtosis of the sum
$w_{CHF} x_{CHF} + w_{JPY} x_{JPY}$ as a function of $w_{CHF}$, denoted $w$ in
the figure, with the ``no-short'' constraint $w_{CHF} + w_{JPY} =1$.
The thick solid line is determined empirically, by direct calculation
of the kurtosis from the data.
The thin solid line
is the theoretical prediction using our theoretical formulas with the
empirically determined exponents $c$ and characteristic scales $\chi$
given above.
While there is a non-negligible difference,
the empirical and theoretical excess kurtosis have essentially the same
behavior with their minimum reached almost at the same value
of $w_{CHF}$.

Three origins of the discrepancy between theory and empirical data
can be invoked. First, as already pointed out in the preceding section,
the modified Weibull distribution with constant exponent and scale parameters
describes accurately only the tail of the empirical distributions while, for
small returns, the empirical distributions are close to a Gaussian law.
While putting a strong emphasis on large fluctuations,
cumulants of order $4$ are still significantly sensitive to the bulk
of the distributions. Moreover, the excess kurtosis is normalized
by the square second-order cumulant, which is almost exclusively
sensitive to the bulk
of the distribution.
Cumulants of higher order should thus be better
described by the modified Weibull distribution. However, a careful
comparison between theory and data would then be hindered by the
difficulty in estimating reliable empirical cumulants of high order.
The second possible origin of the discrepancy between theory and data
is the existence of a weak asymmetry of the empirical distributions,
particularly of the Swiss franc, which has not been taken into account.
The figure also suggests that an error in the determination of the
exponents $c$
can also contribute to the discrepancy.

In order to study investigate the sensitivity with respect to the choice of
the parameters $q$ and $\rho$, we have also constructed the
dashed line corresponding to the theoretical curve with $\rho=0$
(instead of $\rho=0.43$)
and the dotted line corresponding to the
theoretical curve with $q_{CHF}=2$ rather than $1.75$. Finally,
the dashed-dotted line corresponds to the
theoretical curve with $q_{CHF}=1.5$.
We observe that the dashed line remains rather close to the
thin solid line while the dotted line departs significantly when $w$ increases.
Therefore, the most sensitive parameter is $q$, which is natural
because it controls directly the extend of the fat tail of the
distributions.

\begin{figure}
\begin{center}
\includegraphics[width=15cm]{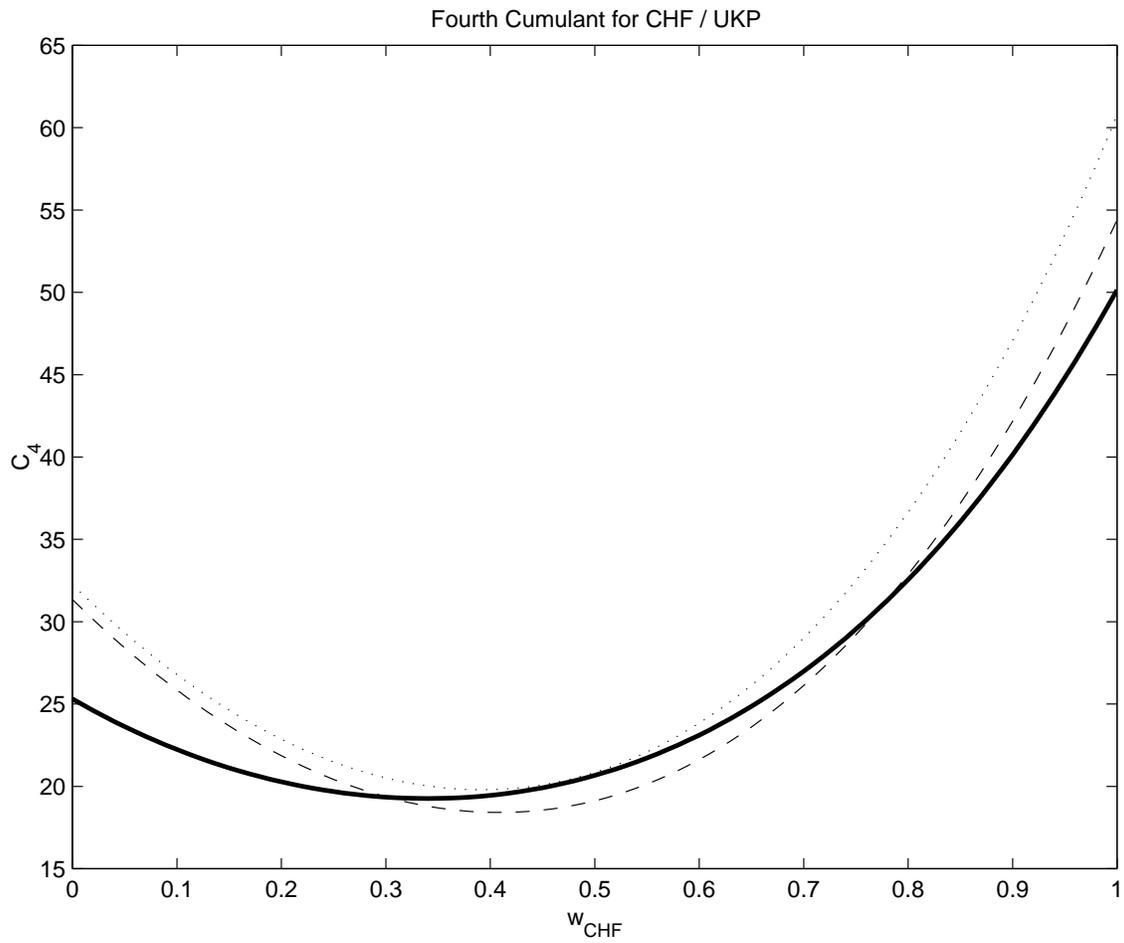}
\end{center}
\caption{\label{fig:c4} Fourth cumulant for a portfolio made of a fraction
$w_{CHF}$ of Swiss Franc and $1-w_{CHF}$ of British Pound. The thick solid line
represents the empirical cumulant while the dotted line represents the
theoretical cumulant under the symmetric assumption. The dashed line shows the
theoretical cumulant when the
slight asymmetry of the assets has been taken into account. This
cumulant has been evaluated
for the time interval from Jan. 1971 to Oct. 1998.}
\end{figure}

In order to account for the effect of asymmetry, we have plotted the fourth
cumulant of a portfolio compound of Swiss Franc and British Pound. On figure
\ref{fig:c4}, the solid line represents the empirical cumulant while the dashed
line shows the theorical cumulant. The agreement between the two curves is
better than under the symmetric asumption. Note once again that an accurate
determination of the parameters is the key point to obtain a good agreement
between empirical data and theoretical prediction. As we can see in the figure
\ref{fig:c4}, the paramaters of the Swiss Franc seem well adjusted since the
theoretical and empirical cumulants are both very close when $w
\simeq 1$, i.e., when
the Swiss Franc is almost the sole asset in the portfolio, while when $w
\simeq 0$, the theorical cumulant is far from the empirical one, i.e., the
parameters of the Bristish Pound are not sufficiently well-adjusted.

\section{Portfolio optimization}

Up to now, we have calculated several measures of risk: the asymptotic
distribution of losses and the corresponding
 Value-at-Risk to quantify the large fluctuations and the cumulants
to evaluate smaller fluctuations. We now present different strategies of
portfolio allocation constructed using these risk measures.

\subsection{Optimization with respect to the Value-at-Risk}

As already said, in the case of assets with different exponent $c$, there
is no possible
diversification of large risks. Therefore, the strategy minimizing large
risks consists in holding as little
as possible of the more fat-tailed assets, compatible with the constraint
on the average return of the portfolio. Obviously, if only large risk
control matters,
one should only hold the asset with the thinnest tail.

The more interesting case occurs when the exponents $c$ are the same for the
different assets of the portfolio. According to
the tables \ref{tab:1} and \ref{tab:2}, this is indeed
an relevant situation, when taking into account the error bars in the
exponents.
It is then easy to show, from the results given
in section \ref{sec:var}, that the VaR is an increasing function of the
scale parameter $\chi$ where:
\be
\chi^c = \frac{1} {c\sum_{i,j} V_{ij}^{-1} \sigma_i^{c/2} \sigma_j^{c/2}}
\ee
when $c>1$ and
\be
\chi^2=\max_{i\in \{1,..N\}} \{w_1^2 \chi_1^2, \cdots, w_N^2 \chi_N^2\}
\ee
if $c\le1$.

Thus, the optimization program to fulfill is~:
\be
\left\{
\begin{array}{l}
{\chi^*} \equiv \inf_{\{w_i\}} {\rm VaR} (\{w_i\})\\
\sum_i w_i = 1 \\
\sum_i w_i \mu_i = \mu
\end{array}
\right.
\ee
where $\mu_i$ denotes the average return of asset $i$ and $\mu$ the average
return of the portfolio.

\paragraph{Case $c>1$ :}
In the case, we can give the analytical form of the efficient
frontier, i.e., $\mu$ versus ${\chi^*}^c$ (at least) in the
independent case. Let
$\lambda_1$ and $\lambda_2$ denote two Langrange multipliers. We have to solve
\be
\frac{\partial}{\partial w_k} \left( \chi - \lambda_1\left( \sum_i w_i
-1 \right) - \lambda_2 \left( \sum_i w_i \mu_i \right) \right)=0~,
\ee
which leads to
\be
\label{eq:w}
\left( \sum_i (w_i^* \chi_i)^\frac{c}{c-1} \right) ^{\frac{c-1}{c}-1}
\chi_k^\frac{c}{c-1} {w_k^*}^{\frac{c}{c-1}-1} = \lambda_1 + \lambda_2 \mu_k~.
\ee

Multiplying by $w_k^*$ and summing over $k$, we obtain~:
\be
\lambda_1+\lambda_2 \mu=\chi~.
\ee
Then, expressing $w_k^*$ from eq.(\ref{eq:w}) and accounting for the
constraint,
we obtain an analytical parametric equation of the efficient frontier :
\be
\left\{
\begin{array}{l}
\sum_i \frac{(\lambda_1+\lambda_2 \mu_i)^{c-1}}{\chi_i^c} = \chi^*\\
\sum_i \frac{\mu_i(\lambda_1+\lambda_2 \mu_i)^{c-1}}{\chi_i^c} =\mu \chi^*\\
\lambda_1+\lambda_2 \mu=\chi^*
\end{array}
\right.
\ee
Varying $\lambda_1$ and $\lambda_2$, the efficient frontier is delineated.
In the Gaussian case $c=2$, we retrieve the standard Markovitz efficient
frontier.

In the case of correlated assets, the shape of the frontier remains essentially
the same but its determination requires numerical calculation.

\paragraph{Case $c<1$ :} In this case, the general minimization problem
must be again solve numerically. As an example, we consider the
simple case where
the portfolio is made of only two assets. This case is analyticaly soluble.
Indeed, the program
\be
\left\{
\begin{array}{l}
\inf_{w^* \in[0,1]} \{\sup_{i\in \{1,..N\}} \{w_1^2 \chi_1^2, \cdots, w_N^2
\chi_N^2\} \}\\
\sum_i w_i = 1 \\
\sum_i w_i \mu_i = \mu
\end{array}
\right.
\ee
simplifies into~:
 \be
\inf_{\mu \in[\mu_1,\mu_2]} \left\{\sup
\left\{\left(\frac{\mu-\mu_2}{\mu_1-\mu_2}\right)^2 \chi_1^2,
\left(\frac{\mu_1-\mu}{\mu_1-\mu_2} \right)^2 \chi_2^2\right\} \right\} \; .
\ee

\begin{figure}
\begin{center}
\includegraphics[width=15cm]{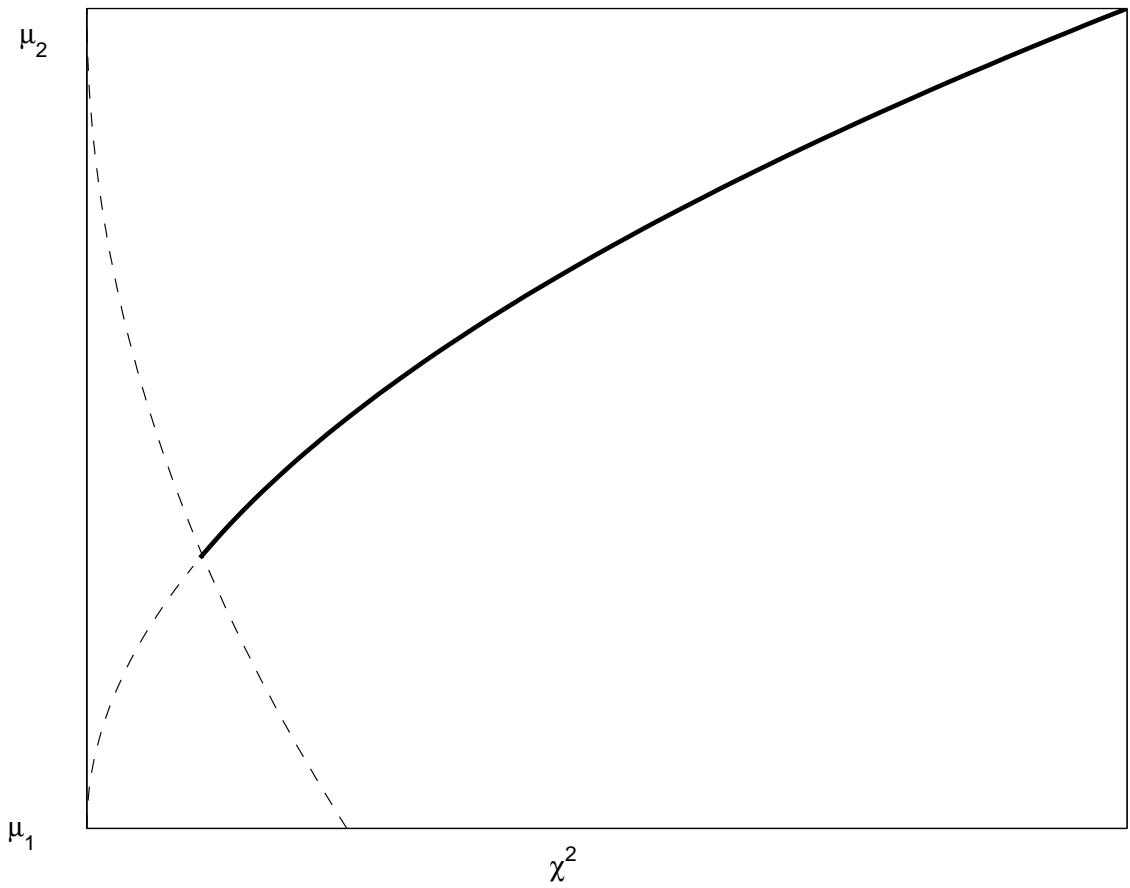}
\end{center}
\caption{\label{fig:var_stretch} Efficient frontier (thick line) in
  the case of two assets with the same exponent lower than $1$.}
\end{figure}

The thick line on figure \ref{fig:var_stretch} represents the efficient
frontier while the dashed lines represent $\mu=f(\chi_1)$ and $\mu=f(\chi_2)$.


\subsection{Optimization with respect to the cumulants}


{\it A priori}, all order cumulants can be considered
and each of them embodies a certain measure of risk. Obviously, the larger
the cumulant order, the more sensitive is the cumulant to large fluctuations.
We can expect risk control based on the high-order cumulants will be
equivalent to
risk-control based on the VaR
approach \cite{Sornette1}. In this section, we will only focus on
relatively low-order cumulants, because they are expected to add
valuable information
on top of the risk structure already provided by the VaR approach.
By lower-order, we mean cumulants of order $n=2$ to $n=6$. To justify
this approach, we have
evaluated the cumulants for several portfolios and it is very clear
that, in most cases for
$n > 6$, they become strictly monotonous functions of the asset weights within
the interval $[0,1]^N$. Therefore, the
minimization of risks which would correspond to minimizing cumulants
of order larger than
$n=6$ would lead once more to hold as little as possible of the more fat-tailed
assets.

The portfolio optimization along the lines of
\cite{Sornette1,SorAnderSim2,AndersenRisk} corresponds to minimizing a given
cumulant $C_n$ of order $n$ of the distribution of returns of the portfolio
in the presence of a constraint on the return as well as the
``no-short'' constraint:
\be
\left\{
\begin{array}{l}
\inf_{w_i\in[0,1]} C_n(\{w_i\})\\
\sum_i w_i = 1 \\
\sum_i w_i \mu_i = \mu~.
\end{array}
\right.
\ee
In the most general case, we have to use extentions for $N > 2$ of the
formulae given in section \ref{sec:cum} and perform a numerical optimization.
Interesting results can be observed in simpler situations, as we
now obtain.

\begin{figure}
\begin{center}
\includegraphics[width=15cm]{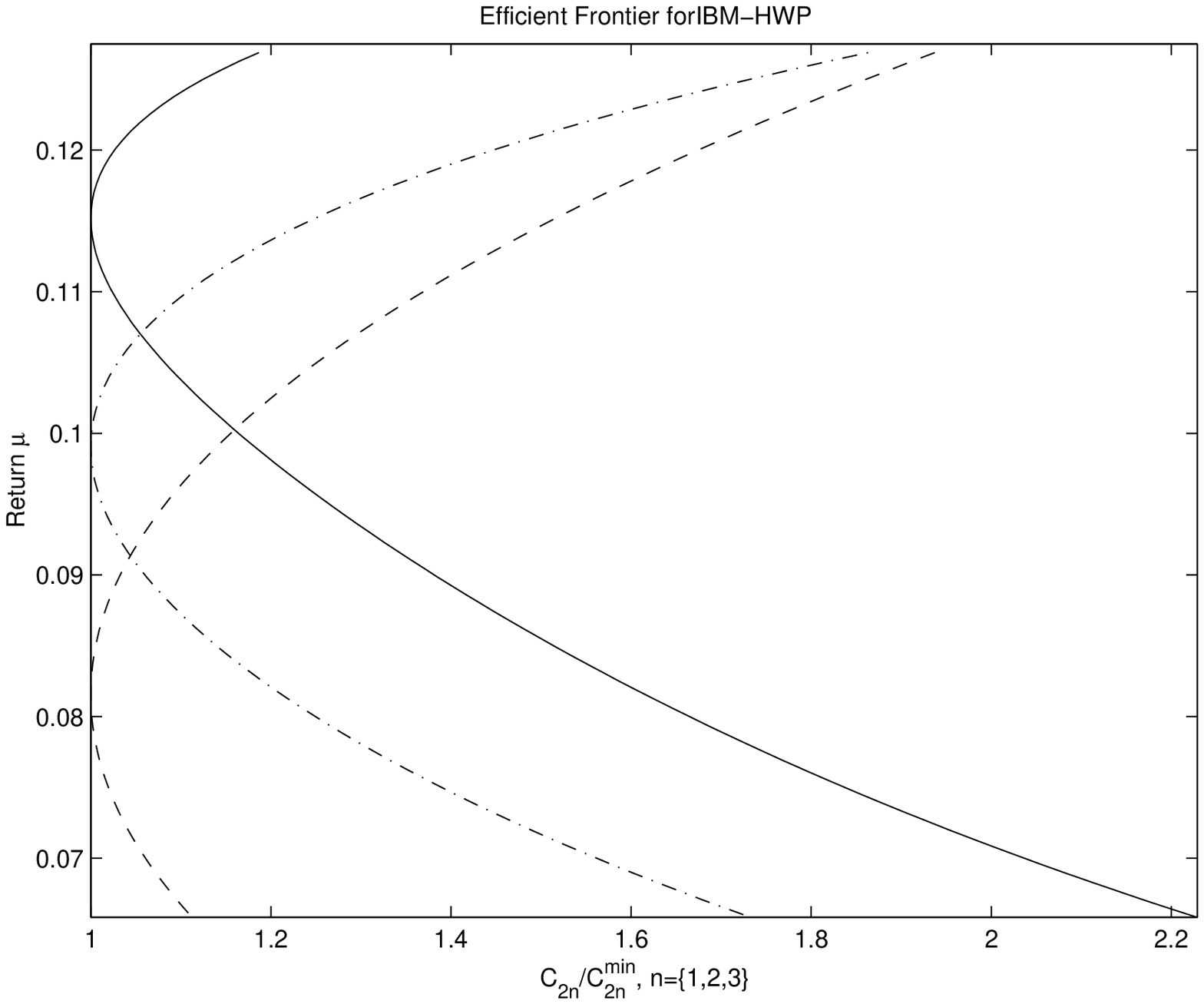}
\end{center}
\caption{\label{fig:ef1} Efficient frontier for a portfolio composed of two
stocks: IBM and Hewlett-Packard. The dashed line represents the efficient
frontier with respect to the second cumulant, i.e., the standard Markovitz
efficient frontier, the dash-doted line represents the efficient frontier
with respect to the fourth cumulant and the solid line the efficient frontier
with respect to the sixth cumulant. The data set used covers the time interval
from Jan. 1977 to Dec 2000.}
\end{figure}

\begin{figure}
\begin{center}
\includegraphics[width=15cm]{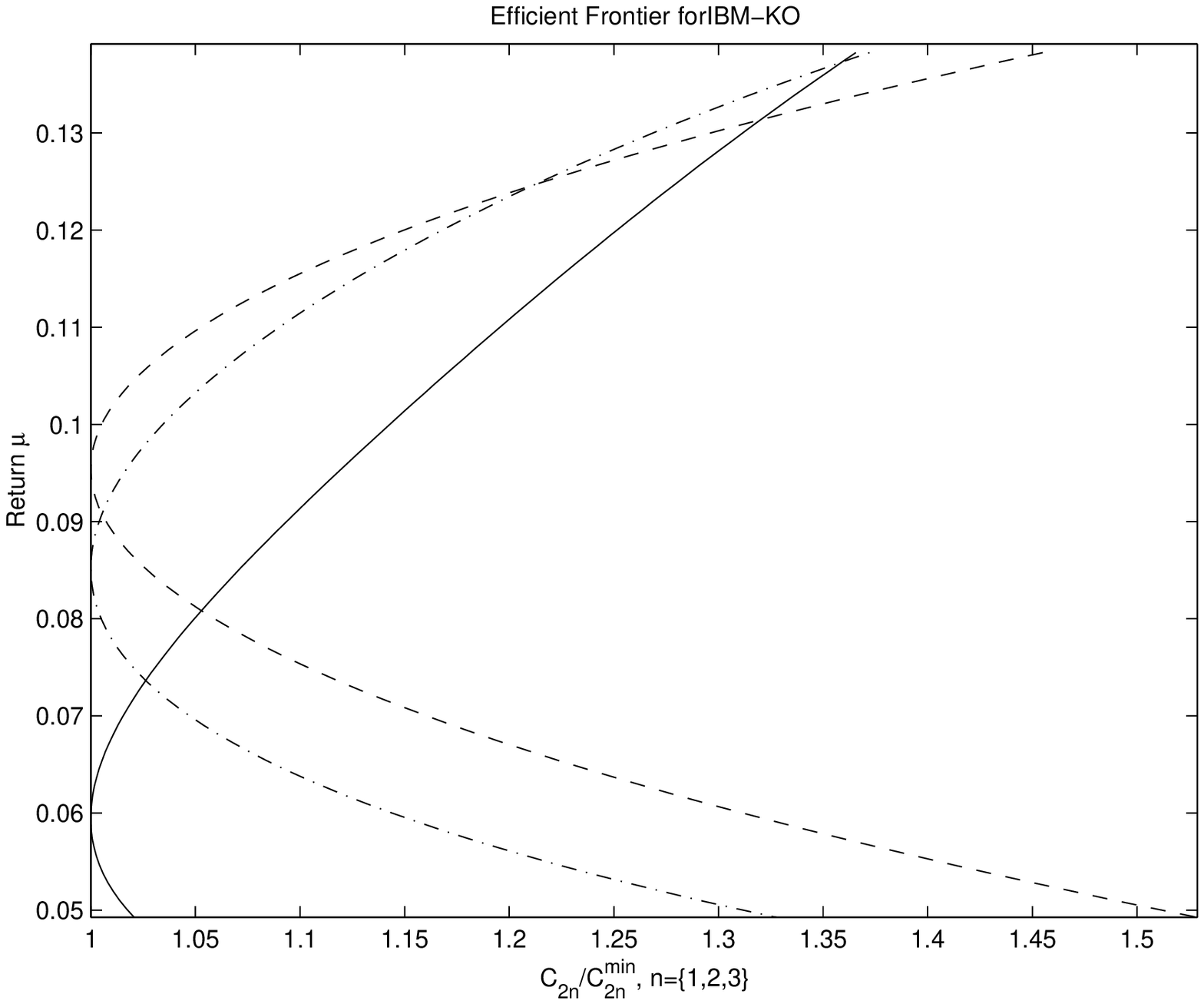}
\end{center}
\caption{\label{fig:ef2} Efficient frontier for a portfolio composed of two
stocks: IBM and Coca-Cola. The dashed line represents the efficient
frontier with respect to the second cumulant,i.e., the standard Markovitz
efficient frontier, the dash-doted line represents the efficient frontier
with respect to the fourth cumulant and the solid line the efficient frontier
with repect to the sixth cumulant. The dataset used covers the time interval
from Jan. 1970 to Dec 2000.}
\end{figure}

Figure \ref{fig:ef1} and \ref{fig:ef2} show the generalized efficient frontiers
using $C_2$ (Markovitz case), $C_4$ or $C_6$ as relevant measures of risks,
for two portfolios composed of two stocks : IBM and Hewlett-Packard in the
first case and IBM and Coca-Cola in the second case.

Obviously, given a certain amount of risk, the mean return of the portfolio
changes when the cumulant considered changes. It is interesting to note that,
in figure \ref{fig:ef1}, the minimisation of larger risks, i.e., with
respect to $C_6$, increases the average return  while, in figure
\ref{fig:ef2}, the minimisation of larger risk lead to decrease the average
return.

This allows us to make precise and quantitative the previously reported
empirical observation that it is possible to ``have your cake and eat it too''
\cite{AndersenRisk}.
We can indeed give a general criterion to determine under which values of the
parameters (exponents $c$ and characteristic scales $\chi$ of the distributions
of the asset returns) the average return of the
portfolio may increase while the large risks decrease {\it at the
same time}, thus
allowing one to gain on both account (of course, the small risks
quantified by the variance
will then increase).
For two independent
assets, assuming that the cumulants of order $n$ and $n+k$ of the portfolio
admit a minimum in the interval $]0,1[$, we can show that
\be
\mu_n^* < \mu_{n+k}^*
\ee
if and only if
\begin{itemize}
\item $\mu_1 - \mu_2 >0$ and
\be
\left( \frac{C_n(1)}{C_n(2)} \right)^{\frac{1}{n-1}} >
\left( \frac{C_{n+k}(1)}{C_{n+k}(2)} \right)^{\frac{1}{n+k-1}},
\ee
\item $\mu_1 - \mu_2 <0$ and
\be
\left( \frac{C_n(1)}{C_n(2)} \right)^{\frac{1}{n-1}} <
\left( \frac{C_{n+k}(1)}{C_{n+k}(2)} \right)^{\frac{1}{n+k-1}},
\ee
\end{itemize}
where $\mu_n^*$ denotes the return of the portfolio evaluated with respect to
the minimum of the cumulant of order $n$ and $C_n(i)$ is the cumulant of order
$n$ for the asset $i$.

The proof of this result and its generalisation to $N>2$ are given in appendix
\ref{app:C}.  In fact, we have observed that when the exponent $c$ of the
assets remains sufficiently different, this result still holds in presence of
dependence between assets.

For the assets considered above, we have found $\mu_{IBM} = 0.13$,
$\mu_{HWP}=0.07$, $\mu_{KO}=0.05$ and
\bea
\frac{C_2(IBM)}{C_2(HWP)} &= 1.76 >
\left(\frac{C_4(IBM)}{C_4(HWP)}\right)^\frac{1}{3} &= 1.03
> \left( \frac{C_6(IBM)}{C_6(HWP)}\right)^\frac{1}{5} = 0.89 \\
\frac{C_2(IBM)}{C_2(KO)} &= 0.96 <
\left(\frac{C_4(IBM)}{C_4(KO)}\right)^\frac{1}{3} &= 1.01
< \left( \frac{C_6(IBM)}{C_6(KO)}\right)^\frac{1}{5} = 1.06~,
\eea
which shows that, for the portfolio IBM / Hewlett-Packard, the
efficient return is
an increasing function of the order of the cumulants while, for the portfolio
IBM / Coca-Cola, the inverse phenomeon occurs. This is exactly what is shown on
figures \ref{fig:ef1} and \ref{fig:ef2}.

The underlying intuitive mechanism is the following: if a portfolio
contains an asset with a rather fat tail (many ``large''
risks) but narrow waist (few ``small'' risks) with very little return to
gain from it, minimizing the variance $C_2$ of the return portfolio
will overweight
this asset which is wrongly perceived as having little risk due to its small
variance (small waist). In contrast, controlling for the larger risks
quantified by $C_4$ or $C_6$ leads to decrease the
weighing of this asset on the portfoio, and correspondingly to
increase the weight of the more
profitable assets. We thus see that the effect of ``both decreasing
large risks and increasing profit'' appears when the asset(s) with
the fatter tails,
and therefore the narrower central part, has(ve) the smaller overall
return(s). A mean-variance
approach will weight them more than deemed appropriate from a prudential
consideration of large risks and consideration of profits.

\section{Conclusion}

We have presented new analytical results on and empirical
tests of a general framework for a portfolio
theory of non-Gaussian risks with non-linear correlations. We have shown that
the concept of efficient frontiers can be generalized by using other
measures of risks
than the variance (cumulant of order $2$),
for instance the value-at-risk and the cumulants of order larger than $2$.
This work opens several novel interesting avenues for research. One consists
in extending the Gaussian copula assumption, for instance by using
the maximum-entropy
principle with non-extensive Tsallis entropies, known to be the
correct mathematical
information-theoretical representation of power laws. A second line of research
would be to extend the present framework to encompass simultaneously different
time scales $\tau$ in the spirit of \cite{MuzyQF} in the case of a
cascade model
of volatilities.

\newpage

\appendix
\section{Asymptotic expansion of the wealth distribution for
a portfolio made of assets with same exponent $c$}
\label{app:A}
\subsection{Case of independent assets}

 We start with the definition $S = \sum_{i=1}^N w_i x_i$ and the
corresponding equation
for its probability density function :
\be
\label{eq:ps}
P_S(S) \propto \int dx_1 \cdots dx_N ~e^{-\sum_{i=1}^N f_i(x_i)}~
\delta\left( S- \sum_{i=1}^N w_i x_i \right)~,
\ee
where we have used the parameterization
\be
P_i(x_i)=C_i~e^{-f_i(x_i)}~,
\ee
and
\be
f_i(x_i)=\left( \frac{|x_i|}{\chi_i} \right)^c~,~ ~ \mbox{with c} >1 ~.
\ee

All equations in (\ref{eq:ps}) are from $-\infty$ to $+ \infty$. The
delta function
expresses the constraint on the sum. We need the following conditions on the
functions $f_i$
\begin{itemize}
\item[(i)] $f_i(x_i) \rightarrow + \infty$ sufficiently fast to
ensure the normalization
of the pdf's.
\item[(ii)] $f_i''(x_i)>0$ (convexity), where $f''$ denotes the
second derivative of
$f$.
\item[(iii)] $\lim_{x \rightarrow \infty} x^2 f(x) =+ \infty$.
\end{itemize}

Under these assumptions, the leading order expansion of $P_S(S)$ for large
$S$ and finite $N>1$ is obtained by a generalization of Laplace's method
which here amounts to remark that the set of $x_i^*$'s that maximize the
integrand in (\ref{eq:ps}) are solution of
\be
\label{eq:min}
f_i'(x_i^*)= \sigma(S) w_i~,
\ee
where $\sigma(S)$ is nothing but a Lagrange multiplier introduced to
minimize the
expression $\sum_{i=1}^N f_i(x_i)$ under the constraint $\sum_{i=1}^N
w_i x_i=S$.

Expanding $f_i(x_i)$ around $x_i^*$ yields
\be
f_i(x_i)=f_i(x_i^*)+f_i'(x_i^*)h_i+\frac{1}{2}f_i''(x_i^*) h_i^2 +
\frac{1}{6} f_i(x_i^*)^{(3)} h_i^3+ \frac{1}{24} f_i^{(4)}(x_i^*) h_i^4  \cdots
\ee
where the $h_i=x_i-x_i^*$ obey the condition
\be
\label{eq:hi}
\sum_{i=1}^N w_i h_i=0~.
\ee

Taking into account equation (\ref{eq:min}), we obtain
\be
\sum f_i(x_i)=\sum f_i(x_i^*) + \sigma(S)\sum w_i hi + \frac{1}{2}
\sum f_i''(x_i^*) h_i^2 +  \frac{1}{6} \sum f_i^{(3)}(x_i^*) h_i^3 +
\frac{1}{24}
\sum f_i^{(4)}(x_i^*) h_i^4 + \cdots~,
\ee
and with (\ref{eq:hi})
\be
\sum f_i(x_i)= \sum f_i(x_i^*) + \frac{1}{2}
\sum f_i''(x_i^*) h_i^2 +  \frac{1}{6} \sum f_i^{(3)}(x_i^*) h_i^3 +
\frac{1}{24}
\sum f_i^{(4)}(x_i^*) h_i^4 + \cdots
\ee

Thus $\exp(- \sum f_i(x_i))$ can be rewritten, up to the
$h_i^4$-order, as follows :
\be
e^{- \sum f_i(x_i)} = e^{ \sum f_i(x_i^*)} e^{ \frac{1}{2}
\sum f_i''(x_i^*) h_i^2}\left[1+ \frac{1}{6} \sum f_i^{(3)}(x_i^*)
h_i^3 +  \frac{1}{24}
\sum f_i^{(4)}(x_i^*) h_i^4 + \cdots \right]
\ee

Then expression (\ref{eq:ps}) becomes
\bea
P_S(S) \propto e^{- \sum_{i=1}^N f_i(x_i^*)} \int dh_1 \cdots dh_N~
e^{-  \frac{1}{2} \sum_{i=1}^N
f_i''(x_i^*) h_i^2} ~\delta\left( \sum_{i=1}^N w_i h_i
\right) \times  \nonumber \\
\times \left[1+ \frac{1}{6} \sum f_i^{(3)}(x_i^*) h_i^3 +  \frac{1}{24}
\sum f_i^{(4)}(x_i^*) h_i^4 + \cdots \right]~.
\eea

Using the fact that
\be
\delta \left( \sum_{i=1}^N w_i h_i \right) = \int \frac{dk}{2 \pi}
~e^{-ik \left( \sum_{j=1}^N w_j h_j \right)}~,
\ee
we obtain
\bea
P_S(S)\propto e^{- \sum_{i=1}^N f_i(x_i^*)} \int \frac{dk}{2 \pi}
\int dh_1 \cdots dh_N~
e^{- \frac{1}{2} \sum_{j=1}^N \left(
f_j''(x_j^*) h_i^2 +2ik w_j h_j \right)} \times  \nonumber \\
\times \left[1+ \frac{1}{6} \sum f_i^{(3)}(x_i^*) h_i^3 +  \frac{1}{24}
\sum f_i^{(4)}(x_i^*) h_i^4 + \cdots \right]~,
\eea
and
\bea
P_S(S) \propto e^{- \sum_{i=1}^N f_i(x_i^*)} \int \frac{dk}{2 \pi}~
e^{-\frac{1}{2} k^2 \sum_{i=1}^N \frac{w_i^2}{f_i''(x_i^*)}}
\int dh_1 \cdots dh_N~ e^{- \frac{1}{2} \sum_{j=1}^N  f_j''(x_j^*) \left(
h_j +ik \frac{w_j}{f_j''(x_j^*)} \right)^2} \times  \nonumber \\
\times  \left[1+ \frac{1}{6} \sum f_i^{(3)}(x_i^*) h_i^3 +  \frac{1}{24}
\sum f_i^{(4)}(x_i^*) h_i^4 + \cdots \right]~.
\eea

Let us denote by $\langle \cdot \rangle_i$ the average over $h_i$ with respect
to a Gaussian distribution whose mean is $-ik
\frac{w_j}{f_j''(x_j^*)}$ and variance : $1/\sqrt{f_i''(x_i^*)}$ and
by $\langle \cdot \rangle_k$ the average over $k$ with respect
to a Gaussian distribution whose mean is $0$ and variance :
$1/\sqrt{\sum \frac{w_i^2}{f_i''(x_i^*)}}$. With these notations, we
can rewrite the equation above as follows~:
\be
P_S(S) \propto \frac{(2 \pi)^{\frac{N-1}{2}}}
{\sqrt{\sum_{i=1}^N \frac{w_i^2 \prod_{j=1}^N f_j''(x_j^*)}{f_i''(x_i^*)}}}
e^{- \sum_{i=1}^N f_i(x_i^*)}
\left[1+ \frac{1}{6} \sum f_i^{(3)}(x_i^*) \langle \langle h_i^3
\rangle_i \rangle_k +  \frac{1}{24}
\sum f_i^{(4)}(x_i^*) \langle \langle h_i^4
\rangle_i \rangle_k + \cdots \right]~.
\ee

The term $\langle h_i^3 \rangle_i$ (and every odd order term in $h_i$)
only involves odd order term in $k$ and thus $\langle \langle h_i^3
\rangle_i \rangle_k$ vanishes, so a simplification occurs :
\be
P_S(S) \propto \frac{(2 \pi)^{\frac{N-1}{2}}}
{\sqrt{\sum_{i=1}^N \frac{w_i^2 \prod_{j=1}^N f_j''(x_j^*)}{f_i''(x_i^*)}}}
e^{- \sum_{i=1}^N f_i(x_i^*)}
\left[1+ \frac{1}{24}
\sum f_i^{(4)}(x_i^*) \langle \langle h_i^4
\rangle_i \rangle_k + \cdots \right]~.
\ee

The evaluation of $ \langle \langle h_i^4 \rangle_i \rangle_k$ leads to
\be
\langle \langle h_i^4 \rangle_i \rangle_k = \frac{3}{f_i''(x_i^*)^2} -
\frac{6 w_i^2}{f_i''(x_i^*)^3 \sum_j \frac{w_j^2}{f_j''(x_j^*)}} +
\frac{3 w_i^4}{f_i''(x_i^*)^4 \left( \sum_j
\frac{w_j^2}{f_j''(x_j^*)}\right)^2}
\ee

Using equation (\ref{eq:min}), we obtain
\be
x_i^*=\left( \frac{\sigma(S)}{c} w_i \chi_i^c \right)^\frac{1}{c-1}~,
\ee
which, together with the constraint $\sum_{i=1}^N w_i x_i^*=S$, gives
\be
x_i^* = \frac{(w_i \chi_i)^{\frac{c}{c-1}}}{w_i \chi^{\frac{c}{c-1}}} S~,
\ee
where
\be
\chi^{\frac{c}{c-1}} = \sum_{i=1}^N (w_i \chi_i)^{\frac{c}{c-1}}.
\ee

Thus, $x_i^* \propto S$ and $f_i^{(n)}(x_i^*) \propto S^{c-n}$. So
$\langle \langle h_i^4 \rangle_i \rangle_k \propto S^{4-2c}$ and
$\sum f_i^{(4)}(x_i^*) \langle \langle h_i^4
\rangle_i \rangle_k \propto S^{-c}$.

Let us now analyze the impact of the higther order terms.
First of all, consider the $6^{th}$ order term
\be
\sum_i f_i^{(6)}(x_i^*) \langle \langle h_i^6 \rangle_i  \rangle_k +
\sum_{i,j} f_i^{(3)}(x_i^*)  f_j^{(3)}(x_j^*)   \langle \langle h_i^3 \rangle_i
\langle h_j^3   \rangle_j  \rangle_k ~,
\ee
where we have omitted the numerical constants.
More generaly, for the $m^{th}$ order term, we will obtain a
contribution of the form
\be
\sum_i f_i^{(m)}(x_i^*) \langle \langle h_i^m \rangle_i  \rangle_k  ~,
\ee
and many other contributions of the form
\be
\sum_{i_1,\cdots,i_n} f_{i_1}^{(p_1)}(x_{i_1}^*) \cdots
f_{i_n}^{(p_n)}(x_{i_n}^*)
\langle \langle h_{i_1}^{p_1} \rangle_{i_1}  \cdots
\langle h_{i_n}^{p_n}   \rangle_{i_n}  \rangle_k ~,
\ee
with $p_1 + \cdots + p_n =m$ and $p_i \geq 3$.

Thus, as a first step, we have to evaluate the $n^{th}$ moment
$\langle h^n \rangle$ of a Gaussian
variable $h$.  Let us denote $m$ and $\sigma$ the mean
and variance of this variable. Using the well known result
\be
\exp \left( -t^2 + 2tx \right) = \sum \frac{t^n}{n!} H_n(x) ~,
\ee
where $H_n(\cdot)$ is the Hermite polynomial of order $n$, we get
\be
\langle h^n \rangle = \frac{1}{i^n} \left(
\frac{\sigma}{\sqrt{2}} \right)^n H_n \left( \frac{im}{\sqrt{2}
\sigma}  \right)~,
\ee
and so
\be
\langle h_j^n \rangle_j = \frac{1}{i^n} \left( \frac{1}{\sqrt{2
f_j''(x_j^*)}} \right)^n
H_n \left( \frac{w_j k}{\sqrt{2 f_j''(x_j^*)}}  \right) ~.
\ee

As a second step, let us evaluate the term $\sum_i f_i^{(2m)}(x_i^*)
\langle \langle h_i^{2m} \rangle_i  \rangle_k$ (recall that
odd order terms vanish).  We have
\bea
\langle \langle h_j^{2m} \rangle_j  \rangle_k   &=&   (-1)^m \int dk
\frac{\sqrt{\sum \frac{w_i^2}{f_i''(x_i^*)}}}{\sqrt{2 \pi}}
\frac{1}{\left( 2 f_j''(x_j^*) \right)^m}  H_{2m} \left( \frac{w_j
k}{\sqrt{2 f_j''(x_j^x)}} \right)
e^{-\frac{k^È}{2} \sum \frac{w_i^2}{f_i''(x_i^*)}}  \nonumber \\
&=& \frac{(-1)^m}{\sqrt{\pi}}  \frac{1}{\left( 2 f_j''(x_j^*)
\right)^m}   \int du ~ e^{-u^2}
H_{2m}  \left( u \sqrt{\frac{w_j^2}{f_j''(x_j^*)}}
\frac{1}{\sqrt{\sum \frac{w_i^2}{f_i''(x_i^*)}}} \right)
\eea
The integral can be calculated exactly but this is not useful as the
only interesting thing is that the
argument in the polynomial $H_{2m}$ is independent of $S$. Indeed, the terms
$f_i''(x_i^*)$'s are proportional to $S^{c-2}$ and cancel out in the
argument of $H_{2m}$. Thus the integral behaves like a constant, and
the dependence with respect to $S$ in $\langle \langle h_j^{2m}
\rangle_j  \rangle_k$  is only given by the prefactor $f_i''(x_i^*)^{-m}$~:
\be
\langle \langle h_j^{2m} \rangle_j  \rangle_k   \propto
\frac{1}{S^{m(c-2)}}  ~.
\ee

As $f_j^{(2m)}(x_j^*)$ behaves like $S^{c-2m}$, we obtain
\be
\sum_i f_i^{(2m)}(x_i^*) \langle \langle h_i^{2m} \rangle_i
\rangle_k \propto \frac{1}{S^{c(m-1)}} ~,
\ee
where $m \ge 2$. Therefore, these terms decrease at least as fast as
$S^{-c}$ (with $c>1$).

The third and final step consists in evaluating the ``mixed''-terms. We have
\be
\langle \langle h_{i_1}^{p_1} \rangle_{i_1}  \cdots
\langle h_{i_n}^{p_n}   \rangle_{i_n}  \rangle_k \propto
\frac{1}{\left( 2 f_{i_1}''(x_{i_1}^*) \right)^{p_1/2}
\cdots \left( 2 f_{i_n}''(x_{i_n}^*) \right)^{p_n/2}}
\langle H_{p_1} \cdots H_{p_n} \rangle_k  ~,
\ee
where, as in the second step, the average of the $H_{p_i}$'s over $k$
is independent of $S$. Thus,
\be
\langle \langle h_{i_1}^{p_1} \rangle_{i_1}  \cdots
\langle h_{i_n}^{p_n}   \rangle_{i_n}  \rangle_k \propto
\frac{1}{S^{(c-2)\frac{p_1+\cdots + p_n}{2}} }
\propto \frac{1}{S^{m(c-2)}}  ~.
\ee

Taking into account the fact that
\be
f_{i_1}^{(p_1)} \cdots f_{i_n}^{(p_n)} \propto S^{c-p_1} \cdots
S^{c-p_n} \propto S^{nc-2m} ~,
\ee
we are led to
\be
\sum_{i_1,\cdots,i_n} f_{i_1}^{(p_1)}(x_{i_1}^*) \cdots
f_{i_n}^{(p_n)}(x_{i_n}^*)
\langle \langle h_{i_1}^{p_1} \rangle_{i_1}  \cdots
\langle h_{i_n}^{p_n}   \rangle_{i_n}  \rangle_k  \propto
\frac{1}{S^{c(m-n)}}~.
\ee

As $p_1 + \cdots + p_n =2m$ and $p_i \ge 3$, we have $3n \le 2m$. So
$m-n \ge \frac{n}{2} \ge 1$ and
the ``mixed''-terms decrease at least like $S^{-c}$.

Finally, we can conclude that
terms of order higher than $1$ are sub-dominant for large $S$ which provides
our exact asymptotic result:
\be
P_S(S) \propto \frac{(2 \pi)^{\frac{N-1}{2}}}
{\sqrt{\sum_{i=1}^N \frac{w_i^2 \prod_{j=1}^N f_j''(x_j^*)}{f_i''(x_i^*)}}}
e^{- \sum_{i=1}^N f_i(x_i^*)} \left( 1 + O\left(\frac{1}{S^c}\right)\right)~.
\ee

>From this, we obtain
\be
f_i(x_i^*)=\frac{(\chi_i w_i)^\frac{c}{c-1}}{\chi^\frac{c^2}{c-1}} S^c~
\Rightarrow~ \sum_{i=1}^N f_i(x_i^*)=\left( \frac{S}{\chi} \right)^c~,
\ee
\be
f_i''(x_i^*)=\frac {c(c-1) w_i^2}{\chi^\frac{c(c-2)}{c-1} (w_i
\chi_i)^c} S^{c-2} \Longrightarrow \prod_{i=1}^N f_i''(x_i^*)=
\left(\frac {c(c-1)}{\chi^\frac{c(c-2)}{c-1}} S^{c-2} \right)^N
\prod_{i=1}^N w_i^{2-c} \chi_i^{-c}
\ee
\be
\sum_{i=1}^N \frac{w_i^2}{f_i''(x_i^*)} = \frac{\chi^2}{c(c-1)}
\left( \frac{S}{\chi} \right)^{2-c}~,
\ee

Finally, we find :
\be
\label{eq:ps1}
P_S(S) \propto \frac{\prod_{i=1}^N
w_i^{\frac{c}{2}-1} \chi_i^\frac{c}{2}}{\chi^{N\frac{2-c}{2(c-1)}-1}}
\left[ \frac{2 \pi}{c(c-1)} \left( \frac{S}{\chi} \right)^{2-c} \right]^{
\frac{N-1}{2}} e^{-\left( \frac{S}{\chi} \right)^c}
\ee

Note that this calculation does not require that the
functions $f_i$'s should be symmetric. Therefore, the result still holds for
asymmetric functions.

More over, it is easy to show that when we consider pdf's whose equations
are given by $g_i(x)~e^{-f_i(x)}$ the same result holds as long as $\ln
g_i(x)$ remains sub-dominant with respect to $f_i(x)$. Therefore, we
just have to
multiply equation (\ref{eq:ps1}) by $\prod g_i(x_i^*)$ to obtain the
corresponding correct result. Thus, for the
pdf's given by (\ref{eq:weibull}), the equation (\ref{eq:ps1})
becomes~:
\be
P(S) \sim \frac{c^\frac{N+1}{2}}{\chi~2^\frac{N+1}{2} (c-1)^\frac{N-1}{2}}
\prod_{i=1}^N (w_i \chi_i) ^{\frac{c-1}{2}+\frac{1}{2(c-1)}} \left(
\frac{|S|}{\chi} \right)^{c-2} e^{-\left( \frac{|S|}{\chi} \right)^c}.
\ee

\subsection{Case of dependent assets}

We assume that the marginal distributions are given by the modified
Weibull distributions:
\be
P_i(x_i)=\frac{1}{2 \sqrt{\pi}}\frac{c}{\chi_i^{c/2}} |x_i|^{c/2-1}
e^{-\left(\frac{|x_i|}{\chi_i} \right)^c}~.
\ee

Under the Gaussian copula assumption, we obtain the following form
for the mutivariate ditribution :
\be
P(x_1, \cdots , x_N) = \frac{c^N}{2^N \pi^{N/2}\sqrt{V}} \prod_{i=1}^N
\frac{x_i^{c/2-1}}{\chi_i^{c/2}} \exp\left[-\sum_{i,j}
V_{ij}^{-1}\left(\frac{x_i}{\chi_i}\right)^{c/2}
\left(\frac{x_j}{\chi_j}\right)^{c/2} \right]~.
\ee

Let
\be
f(x_1, \cdots, x_N) = \sum_{i,j}
V_{ij}^{-1}\left(\frac{x_i}{\chi_i}\right)^{c/2}
\left(\frac{x_j}{\chi_j}\right)^{c/2}~.
\ee

We have to maximize $f$ under the constraint $\sum w_i x_i = s$. As
for the independent case,
we introduce a Lagrange multiplier $\lambda$ which leads to
\be
c\sum_{j}
V_{jk}^{-1}\left(\frac{x_j^*}{\chi_j}\right)^{c/2}
\left(\frac{x_k^*}{\chi_k}\right)^{c/2}= \lambda w_k  x_k^*~,
\ee
and using the constraint $\sum w_i x_i^* =S$~:
\be
c\sum_{j,k}
V_{jk}^{-1}\left(\frac{x_j^*}{\chi_j}\right)^{c/2}
\left(\frac{x_k^*}{\chi_k}\right)^{c/2}= \lambda S~.
\ee

Inspired by the results obtained for independent assets and by dimensional
arguments, we assert that $x_i^*$ is proportional to $S$ and $\lambda$
proportional to $S^{c-1}$~:
\bea
x_i^* &=& \chi_i \sigma_i S \\
\lambda &=& K S^{c-1}~,
\eea
where $\sigma_i$ and $K$ depend on $\{w_j\}$ and $\{\chi_j\}$ but are
independent of $S$.

Therefore,
\be
c\sum_{i}
V_{ik}^{-1} \sigma_i^{c/2} \sigma_k^{c/2}= K w_k \chi_k \sigma_k~,
\ee
and
\be
c\sum_{i,j}
V_{ij}^{-1} \sigma_i^{c/2} \sigma_j^{c/2}= K~.
\ee

Thus
\bea
f(x_1, \cdots , x_N) &=& f(x_1^*, \cdots , x_N^*) + \sum_i
\frac{\partial f}{\partial x_i} h_i + \frac{1}{2} \sum_{ij} \frac{\partial^2
f}{\partial x_i \partial x_j} h_i h_j + \cdots \\
&=& c\sum_{j,k} V_{jk}^{-1} \sigma_j^{c/2} \sigma_k^{c/2} S^c +
\frac{1}{2} \sum_{ij} \frac{\partial^2
f}{\partial x_i \partial x_j} h_i h_j + \cdots ~,
\eea
where, as in the previous section, $h_i = x_i-x_i^*$.

It is easy to check that the $n^{th}$-order derivative of $f$ with respect
to the $x_i$'s evaluated at $\{x_i^*\}$ is proportional to $S^{c-n}$.
In the sequel, we
will use the following notation~:
\be
\left. \frac{\partial^n f}{\partial x_{i_1} \cdots \partial x_{i_n}}
\right|_{\{x_i^*\}} = M_{{i_1} \cdots {i_n}}^{(n)} S^{c-n}~.
\ee

We can write :
\be
f(x_1, \cdots , x_N) =  c\sum_{j,k} V_{jk}^{-1} \sigma_j^{c/2}
\sigma_k^{c/2} S^c + \frac{S^{c-2}}{2} \sum_{ij} M_{ij}^{(2)} h_i h_j
+ \frac{S^{c-3}}{6} \sum_{ijk} M_{ijk}^{(3)} h_i h_j h_k + \cdots
\ee
up to the fourth order. This leads to
\bea
P(S)\propto e^{- c\sum_{j,k} V_{jk}^{-1} \sigma_j^{c/2}
\sigma_k^{c/2} S^c} \int dh_1 \cdots dh_N e^{- \frac{S^{c-2}}{2}
\sum_{ij} M_{ij}^{(2)} h_i h_j }~\delta \left( \sum w_i h_i \right)
\times \nonumber \\
\times \left[ 1 + \frac{S^{c-3}}{6} \sum_{ijk} M_{ijk}^{(3)} h_i h_j h_k +
\cdots \right]~.
\eea

Using the same method as in the previous section :
\bea
P(S)\propto e^{- c\sum_{j,k} V_{jk}^{-1} \sigma_j^{c/2}
\sigma_k^{c/2} S^c} \int \frac{dk}{2 \pi} \int dh_1 \cdots dh_N e^{-
\frac{S^{c-2}}{2}
\sum_{ij} M_{ij}^{(2)} h_i h_j -ik \sum_j w_j k_j}
\times \nonumber \\
\times \left[ 1 + \frac{S^{c-3}}{6} \sum_{ijk} M_{ijk}^{(3)} h_i h_j h_k +
\cdots \right]~,
\eea
or in vectorial notation :
\bea
P(S)\propto e^{- c\sum_{j,k} V_{jk}^{-1} \sigma_j^{c/2}
\sigma_k^{c/2} S^c} \int \frac{dk}{2 \pi} \int d{\bf h}~ e^{- \frac{S^{c-2}}{2}
{\bf h^t M^{(2)} h} -ik {\bf w^t h}}
\times \nonumber \\
\times \left[ 1 + \frac{S^{c-3}}{6} \sum_{ijk} M_{ijk}^{(3)} h_i h_j h_k +
\cdots \right]~.
\eea

Let us perform the following standard change of variables~:
\be
{\bf h} = {\bf h'}-\frac{ik}{S^{c-2}} {\bf {M^{(2)}}^{-1} w}~,
\ee
(${\bf {M^{(2)}}^{-1}}$ exists since $f$ is assumed convex and thus
${\bf M^{(2)}}$ positive)~:
\be
\frac{S^{c-2}}{2}{\bf h^t M^{(2)} h} +ik {\bf w^t h} =
\frac{S^{c-2}}{2}{\bf h'^t M^{(2)} h'} + \frac{k^2}{2S^{c-2}} {\bf
w^t  {M^{(2)}}^{-1} w}~.
\ee
This yields
\bea
P(S)\propto e^{- c\sum_{j,k} V_{jk}^{-1} \sigma_j^{c/2}
\sigma_k^{c/2} S^c} \int \frac{dk}{2 \pi}~e^{-\frac{k^2}{2S^{c-2}} {\bf
w^t  {M^{(2)}}^{-1} w}}\times \nonumber \\
\times \int d{\bf h}~ e^{-\frac{S^{c-2}}{2}
\left({\bf h}+\frac{ik}{S^{c-2}} {\bf {M^{(2)}}^{-1} w} \right)^t
 M^{(2)}\left({\bf h}+\frac{ik}{S^{c-2}} {\bf {M^{(2)}}^{-1} w} \right) }
\left[ 1 + \frac{S^{c-3}}{6} \sum_{ijk} M_{ijk}^{(3)} h_i h_j h_k +
\cdots \right]~.
\eea

Denoting by $\langle \cdot \rangle_h$ the average with respect
to the Gaussian distribution of ${\bf h}$ and by $\langle \cdot
\rangle_k$ the average with respect to the Gaussian distribution of
$k$, we have~:
\bea
P(S)\propto \sqrt{ \frac{\det {\bf{M^ {(2)} }^{-1}} } { {\bf w^t {M^
{(2)} }^{-1} w} } }~
(2 \pi S^{2-c})^{\frac{N-1}{2}}~
e^{- c\sum_{j,k} V_{jk}^{-1} \sigma_j^{c/2}\sigma_k^{c/2} S^c} \times
\nonumber \\
\times \left[ 1 + \frac{S^{c-3}}{6} \sum_{ijk} M_{ijk}^{(3)}\langle \langle
h_i h_j h_k \rangle_{\bf h} \rangle_k +  \frac{S^{c-4}}{24}
\sum_{ijkl}  M_{ijkl}^{(4)}\langle \langle
h_i h_j h_k h_l\rangle_{\bf h} \rangle_k + \cdots \right]~.
\eea

We now invoke Wick's theorem, which states that each term $\langle
\langle h_i \cdots
h_p \rangle_{\bf h} \rangle_k$ can be expressed as a product of pairwise
correlation coefficients. Evaluating the average with
respect to the symmetric distribution of $k$, it is obvious that
odd-order terms will vanish
and that the count of powers of $S$ involved in each even-order term
done for independent assets remains true. So, up to the leading
order~:
\be
P(S)\propto \sqrt{ \frac{\det {\bf{M^ {(2)} }^{-1}} } { {\bf w^t {M^
{(2)} }^{-1} w} } }~
(2 \pi S^{2-c})^{\frac{N-1}{2}}~
e^{- c\sum_{j,k} V_{jk}^{-1} \sigma_j^{c/2}\sigma_k^{c/2} S^c}~.
\ee

\newpage

\section{Calculation of the moments of the distribution of portfolio returns}
\label{app:B}

Let us start with equation (\ref{eq2.5}) in the $2$-asset case~:
\bea
\hat P_S(k) = \frac{1}{2 \pi \sqrt{1-\rho^2}} \int dy_1 dy_2~ exp \left[
-\frac{1}{2} y^t V^{-1} y + ik \left( \chi_1 w_1 sgn(y_1)
\left| \frac{y_1}{\sqrt{2}} \right|^{q_1} +  \right. \right. {\nonumber} \\
\left. \left.  + \chi_2 w_2 sgn(y_2) \left| \frac{y_2}{\sqrt{2}}
\right|^{q_2} \right) \right]
\; .
\eea
Expanding the exponential and using the definition
(\ref{eq:defmom}) of moments, we get
\bea
M_n = \frac{1}{2 \pi \sqrt{1-\rho^2}} \int dy_1 dy_2 \sum_{p=0}^n {n \choose
p} \chi_1^p \chi_2^{n-p} w_1^p w_2^{n-p} sgn(y_1)^p \left|
\frac{y_1}{\sqrt{2}} \right|^{q_1p} \times
\nonumber \\
\times sgn(y_2)^{n-p} \left| \frac{y_2}{\sqrt{2}}
\right|^{q_2(n-p)}\, e^{-\frac{1}{2} y^t
V^{-1} y } \; .
\eea
Posing
\be
\label{eq:a1}
\gamma_{q_1 q_2}(n,p) = \frac{ {\chi_1}^p {\chi_2}^{n-p}}{2 \pi
\sqrt{1-\rho^2}} \int dy_1 dy_2 \; sgn(y_1)^p \left| \frac{y_1}{\sqrt{2}}
\right|^{q_1p} sgn(y_2)^{n-p} \left| \frac{y_2}{\sqrt{2}} \right|^{q_2(n-p)}\,
e^{-\frac{1}{2} y^t V^{-1} y } \; ,
\ee
this leads to
\be
M_n = \sum_{p=0}^n {n \choose p} w_1^p w_2^{n-p}
\gamma_{q_1 q_2}(n,p) \; .
\ee

Let us defined the auxiliary variables $\alpha$ and $\beta$ such that
\be
\left\{ \begin{array}{rlc}
\alpha &=& (V^{-1})_{11} = (V^{-1})_{22}  =  \frac{1}{1 - \rho^2} \; , \\
\beta   &=& -(V^{-1})_{12} =-(V^{-1})_{21}  =\frac{\rho}{1 - \rho^2 } \; .
\end{array} \right.
\ee
Performing a simple change of variable in (\ref{eq:a1}), we can transform
the integration such that it is defined solely within the first quadrant
($y_1 \ge 0$, $y_2 \ge 0$), namely
\bea
\label{eq:a2}
\gamma_{q_1 q_2}(n,p) ={\chi_1}^p {\chi_2}^{n-p} \frac{1+(-1)^n}{2 \pi
\sqrt{1-\rho^2}} \int_0^{+\infty} dy_1 dy_2~
\left(\frac{y_1}{\sqrt{2}}\right)^{q_1p} \left(
\frac{y_2}{\sqrt{2}}\right)^{q_2(n-p)}\, e^{-\frac{\alpha}{2}   (y_1^2+y_2^2)}
\times \nonumber \\
\times  \left( e^{\beta y_1 y_2} + (-1)^p e^{-\beta y_1 y_2}
\right) \; . \eea

This equation imposes that the coefficients $\gamma$ vanish if $n$ is odd.
This leads to the vanishing of the moments of odd orders, as expected for
a symmetric distribution. Then, we expand
$e^{\beta y_1 y_2} + (-1)^p e^{-\beta y_1 y_2}$ in series. Permuting
the sum sign and the integral allows us to decouple the integrations
over the two variables $y_1$ and $y_2$:
\bea
\gamma_{q_1 q_2}(n,p) ={\chi_1}^p {\chi_2}^{n-p} \frac{1+(-1)^n}{2 \pi
\sqrt{1-\rho^2}} \sum_{s=0}^{+\infty}[1+(-1)^{p+s}] \frac{\beta^s}{s!} \left(
\int_0^{+\infty} dy_1~ \frac{y_1^{q_1p+s}}{2^\frac{q_1p}{2}} ~
e^{-\frac{\alpha}{2} y_1^2}  \right) \times \nonumber\\
\times \left(
\int_0^{+\infty}
dy_2~\frac{y_2^{q_2(n-p)+s}}{2^\frac{q_2(n-p)}{2}}~e^{-\frac{\alpha}{2}
y_1^2} \right) \; . \eea
This brings us back to the problem of calculating the same type of integrals
as in the uncorrelated case. Using the expressions of $\alpha$ and
$\beta$, and taking into account the parity of $n$ and $p$, we obtain:
\bea
\label{eq:a3}
\gamma_{q_1 q_2}(2n,2p)&=& {\chi_1}^{2p} {\chi_2}^{2n-2p}\frac{(1-\rho^2)^{q_1p
+ q_2(n - p) + \frac{1}{2}}}{\pi}\sum_{s=0}^{+\infty}
\frac{(2\rho)^{2s}}{(2s)!} \Gamma \left( q_1p+s+\frac{1}{2} \right)\times
\nonumber \\
&&\times  \Gamma \left(  q_2(n-p)+s+\frac{1}{2} \right)  \; ,\\
\label{eq:a4}
\gamma_{q_1 q_2}(2n,2p+1)&=&{\chi_1}^{2p+1} {\chi_2}^{2n-2p-1}
\frac{(1-\rho^2)^{q_1p + q_2(n - p) +
\frac{q_1-q_2+1}{2}}}{\pi}\sum_{s=0}^{+\infty}
\frac{(2\rho)^{2s+1}}{(2s+1)!}
\times \nonumber \\
 && \times \Gamma \left( q_1p+s+1+\frac{q_1}{2} \right)  \Gamma
\left(q_2(n-p)+s+1-\frac{q_2}{2} \right) \; . \eea

Using the definition of the hypergeometric functions $_2F_1$
\cite{AB}, and the relation (9.131) of \cite{Gradshteyn}, we finally obtain
\bea
\gamma_{q_1 q_2}(2n,2p) &=&  {\chi_1}^{2p} {\chi_2}^{2n-2p} \frac{\Gamma
\left( q_1p +\frac{1}{2} \right) \Gamma \left(q_2(n-p)+\frac{1}{2}
\right)}{\pi}~
{_2F_1} \left(-q_1p, -q_2(n-p) ; \frac{1}{2} ; \rho^2 \right) \; , \\
\gamma_{q_1 q_2}(2n,2p+1) &=&{\chi_1}^{2p+1} {\chi_2}^{2n-2p-1} \frac{2\Gamma
\left( q_1p+1+\frac{q_1}{2}   \right) \Gamma \left(q_2(n-p)+1-\frac{q_2}{2}
\right)}{\pi} \rho~ \times \nonumber \\
&&\times~ {_2F_1}\left( -q_1p - \frac{q_1-1}{2},  -q_2(n-p)+ \frac{q_2+1}{2} ;
\frac{3}{2} ;\rho^2 \right) \; .
\eea

In the asymmetric case, a similar calculation follows, with the sole
difference that the results
involves four terms in the integral (\ref{eq:a2}) instead of two.

\newpage

\section{Conditions under which it is possible to increase the return
and decrease large risks simultaneously}
\label{app:C}

We consider $N$ independent assets $\{1 \cdots N\}$, whose returns are denoted
by $\mu_1 \cdots \mu_N$. We aggregate these assets in a portfolio. Let $w_1
\cdots w_N$ be their weights. We consider that short positions
are forbiden and that $\sum_i w_i=1$. The return $\mu$ of the portfolio is
\be
\mu = \sum_{i=1}^N w_i \mu_i.
\ee
The risk of the portfolio is quantized by the cumulants of the distribution of
$\mu$.

Let us denote $\mu_n^*$ the return of the portfolio evaluated for asset weights
which minimize the cumulant of order $n$.

\subsection{Case of two assets}

Let $C_n$ be the cumulant of order n for the portfolio. The assets
being independent,
we have
\bea
C_n &=& C_n(1){w_1}^n + C_n(2) {w_2}^n ,\\
&=& C_n(1){w_1}^n + C_n(2) (1-w_1)^n .
\eea

In the following, we will drop the subscript $1$ in $w_1$, and only write $w$.
Let us evaluate the value $w=w^*$ at the minimum of $C_n$, $n>2$ :
\bea
\frac{d C_n}{d w} = 0 &\Longleftrightarrow& C_n(1) w^{n-1} - C_n(2)
(1-w)^{n-1}=0 ,\\
&\Longleftrightarrow& \frac{C_n(1)}{C_n(2)} = \left(
\frac{1-w^*}{w^*} \right)^{n-1}, \eea
and assuming that $C_n(1)/C_n(2)>0$, we obtain
\be
w^* =
\frac{C_n(2)^{\frac{1}{n-1}}}{C_n(1)^{\frac{1}{n-1}}+C_n(2)^{\frac{1}{
n-1}}}.
\ee

This leads to the following expression for $\mu_n^*$ :
\be
\mu_n^* = \frac{\mu_1 \cdot C_n(2)^\frac{1}{n-1} + \mu_2 \cdot
C_n(1)^\frac{1}{n-1}}
{C_n(1)^\frac{1}{n-1} + C_n(2)^\frac{1}{n-1}}.
\ee

Thus, after simple algebraic manipulations, we find
\be
\mu_n^* < \mu_{n+k}^* \Longleftrightarrow (\mu_1-\mu_2)
\left( C_n(1)^\frac{1}{n-1} C_{n+k}(2)^\frac{1}{n+k-1}
- C_n(2)^\frac{1}{n-1} C_{n+k}(1)^\frac{1}{n+k-1} \right) >0,
\ee
which concludes the proof of the result annouced in the main body of the text.

\subsection{General case}
We consider a portfolio with $N$ independent assets.
Assuming that the cumulants $C_n(i)$ have the same sign for all $i$,
the minimum of $C_n$ is
obtained for a portfolio whose weights are given by
\be
w_i = \frac{\prod_{j \neq i}^N C_n(j)^\frac{1}{n-1}}
{\sum_{j=1}^N C_n(j)^\frac{1}{n-1}} ,
\ee
and we have
\be
\mu_n^* = \frac{\sum_{i=1}^N \left(
\mu_i \prod_{j \neq i}^N C_n(j)^\frac{1}{n-1} \right)}
{\sum_{j=1}^N C_n(j)^\frac{1}{n-1}}~.  \label{njfkalkaa}
\ee

Indeed, the following conditions hold:
\bea
\sum_{i=1}^N C_n(i) w_i^{n-1} dx_i = 0 ,\\
\sum_{i=1}^N dx_i = 0 .
\eea

Introducing a Lagrange multiplier $\lambda$, we obtain
\be
w_i^{n-1} = -\frac{\lambda}{C_n(i)}.
\ee

Assuming that all the $C_n(i)$ have the same sign, we can find a $\lambda$
such that all the $w_i$ are real and positive :
\be
w_i = \frac{\prod_{j \neq i}^N C_n(j)^\frac{1}{n-1}}
{\sum_{j=1}^N C_n(j)^\frac{1}{n-1}} .
\ee

If some $C_n(i)$ are positive and others negative, the set of assets we
consider is not compatible with a global minimum on $]0,1[^N$. Then,
we have to split the
portfolio into two sub-portfolios constituted of assets whose cumulants
have the same sign and perform the minimization of the corresponding
cumulant of each subset.

There is not simple condition that ensure $\mu_n^* <\mu_{n+k}^*$. The
simplest way to compare $\mu_n^*$ and $\mu_{n+k}^*$ is to
calculate diretly these quantities using the formula (\ref{njfkalkaa}).

\newpage

\begin{table}[!h]
\begin{center}
\begin{tabular} {|c||c|c|c|c||c|c|c|c|}
\hline
&       $<\chi_+>$&     $<c_+>$&  $\chi_+$&     $c_+$&  $<\chi_->$&
$<c_->$&    $\chi_-$& $c_-$\\
\hline
CHF&    2.45&   1.61&   2.33&   1.26&   2.34&   1.53&   1.72&   0.93\\
\hline
DEM&    2.09&   1.65&   1.74&   1.03&   2.01&   1.58&   1.45&   0.91\\
\hline
JPY&    2.10&   1.28&   1.30&   0.76&   1.89&   1.47&   0.99&   0.76\\
\hline
MAL&    1.00&   1.22&   1.25&   0.41&   1.01&   1.25&   0.44&   0.48\\
\hline
POL&    1.55&   1.02&   1.30&   0.73&   1.60&   2.13&   1.25&   0.62\\
\hline
THA&    0.78&   0.75&   0.75&  0.54&    0.82&   0.73&   0.30& 0.38\\
\hline
UKP&    1.89&   1.52&   1.38&   0.92&   2.00&   1.41&   1.82&   1.09\\
\hline
\end{tabular}
\end{center}
\caption{\label{tab:1} Table of the exponents $c$ and the scale
parameters $\chi$ for different currencies. The subscript ''+'' or
''-'' denotes the positive or negative part of the distribution of returns and
the terms between brackets refer to parameters estimated in the bulk
of the distribution while naked parameters refer to the tails of the
distribution.}
\end{table}

\begin{table}[!h]
\begin{center}
\begin{tabular} {|c||c|c|c|c||c|c|c|c|}
\hline
&       $<\chi_+>$&     $<c_+>$&  $\chi_+$&     $c_+$&  $<\chi_->$&
$<c_->$&    $\chi_-$& $c_-$\\
\hline
AMAT&   12.47&   1.82&   8.75&   0.99&   11.94&   1.66&   8.11
&   0.98 \\
\hline
C&      6.54&   1.70&   2.34&   0.66&   5.99&   1.70&   0.40& 0.36 \\
\hline
EMC&    13.53&   1.63&   13.18&   1.55&   11.44&   1.61&   3.05&   0.57  \\
\hline
F&      7.37&   1.52&   8.35&   1.64&   6.09&   1.91&   5.97&  1.34\\
\hline
GE&     5.21&   1.89&   1.81&   1.28&   4.80&   1.81&   4.31&   1.16\\
\hline
GM&     5.78&   1.71&   0.63&   0.48&   5.32&   1.89&   2.80&   0.79\\
\hline
HWP&    7.51&   1.93&   4.20&   0.84&   7.26&   1.76&   1.66&   0.52\\
\hline
IBM&    5.46&   1.71&   3.85&   0.87&   5.07&   1.90&   0.18&   0.33\\
\hline
INTC&   8.93&   2.31&   2.79&   0.64&   9.14&   1.60&   3.56&   0.62\\
\hline
KO&     5.38&   1.88&   4.46&   1.04&   5.06&   1.74&   2.98&   0.78\\
\hline
LU&     10.46&  2.02&   7.12&   0.98&   10.16&  1.79&   2.11&   0.43\\
\hline
MDT&    6.82&   1.95&   6.09&   1.11&   6.49&   1.54&   2.55&   0.67\\
\hline
MRK&    5.36&   1.91&   4.56&   1.16&   5.00&   1.73&   1.32&   0.59\\
\hline
MSFT&   8.14&   2.19&   2.11&   0.58&   7.77&   1.60&   0.67&   0.38\\
\hline
PFE&    6.41&   2.01&   5.84&   1.27&   6.04&   1.70&   0.26&   0.35\\
\hline
PG&     4.86&   1.83&   3.53&   0.96&   4.55&   1.74&   2.96&   0.82\\
\hline
QCOM&   15.15&  1.70&   14.76&  1.40&   13.02&  1.86&   10.17&  1.07\\
\hline
SBC&    5.21&   1.97&   1.26&   0.59&   4.89&   1.59&   1.56&   0.60\\
\hline
SUNW&   11.54&  1.94&   6.91&   0.90&   10.96&  1.70&   5.89&   0.77\\
\hline
T&      5.13&   1.48&   2.86&   0.76&   4.60&   1.70&   1.87&   0.56\\
\hline
TXN&    9.06&   1.78&   4.07&   0.72&   8.24&   1.84&   2.18&   0.54\\
\hline
WCOM&9.80&      1.74&   11.01&  1.56&   9.09&   1.56&   2.86&   0.58\\
\hline
WMT&    7.41&   1.83&   5.81&   1.01&   6.80&   1.64&   3.75&   0.78\\
\hline
\end{tabular}
\end{center}
\caption{\label{tab:2} Table of the exponents $c$ and the scale
parameters $\chi$ for different stocks. The subscript ''+'' or
''-'' denotes the positive or negative part of the distribution and
the terms between brackets refer to parameters estimated in the bulk
of the distribution while naked parameters refer to the tails of the
distribution.}
\end{table}

\newpage

\newpage

\newpage

\newpage

\newpage

\newpage

\newpage

\newpage

\newpage

\newpage

\newpage

\newpage

\newpage

\newpage

\newpage

\newpage

\newpage

\newpage

\end{document}